\documentclass[pre,aps,onecolumn,superscriptaddress,notitlepage,nofootinbib]{revtex4-1}
\usepackage{amssymb,epsfig,amsmath,bm,subfigure,graphicx,textcomp,url,float,color,cases}
\usepackage{epsfig}
\usepackage{textcomp}
\usepackage[normalem]{ulem}

\usepackage[colorlinks=true, urlcolor=blue, anchorcolor=blue, citecolor=blue,filecolor=blue,linkcolor=blue,menucolor=blue%,pagecolor=blue
]{hyperref}

\usepackage{color}

\usepackage[titletoc,title]{appendix}

\begin{document}
\title{Dynamical large deviations of the fractional Ornstein-Uhlenbeck process}

\author{Alexander Valov}
\email{aleksandr.valov@mail.huji.ac.il}
\affiliation{Racah Institute of Physics, Hebrew University of Jerusalem, Jerusalem 91904, Israel}

\author{Baruch Meerson}
\email{meerson@mail.huji.ac.il}
\affiliation{Racah Institute of Physics, Hebrew University of Jerusalem, Jerusalem 91904, Israel}

\begin{abstract}
The fractional Ornstein-Uhleneck (fOU) process is described by the overdamped Langevin equation $\dot{x}(t)+\gamma x=\sqrt{2 D}\xi(t)$, where $\xi(t)$ is the fractional Gaussian noise with the Hurst exponent $0<H<1$. For $H\neq 1/2$ the fOU process is non-Markovian but Gaussian, and it has either vanishing (for $H<1/2$), or divergent (for $H>1/2$) spectral density at zero frequency. For $H>1/2$, the fOU is long-correlated. Here we study dynamical large deviations of the fOU process and focus on the  area $A_n=\int_{-T}^{T} x^n(t) dt$, $n=1,2,\ldots$ over a long time window $2T$.  Employing the optimal fluctuation method,  we determine the optimal path of the conditioned process, which dominates the large-$A_n$ tail of the probability distribution of the area, $\mathcal{P}(A_n,T)\sim \exp[-S(A_n,T)]$. We uncover a nontrivial phase diagram of scaling behaviors of the optimal paths and of the action $S(A_n\equiv 2 a_n T,T)\sim  T^{\alpha(H,n)} a^{2/n}_n$  on the $(H,n)$ plane. The phase diagram includes three regions: (i) $H>1-1/n$, where $\alpha(H,n)=2-2H$, and the optimal paths are delocalized, (ii)  $n=2$ and $H\leq \frac{1}{2}$, where $\alpha(H,n)=1$, and the optimal paths oscillate with an $H$-dependent frequency, and (iii) $H\leq 1-1/n$ and $n>2$, where $\alpha(H,n)=2/n$, and the optimal paths are strongly localized.  We verify our theoretical predictions in large-deviation simulations of the fOU process. By combining the Wang-Landau Monte-Carlo algorithm with the circulant embedding method of generation of stationary Gaussian fields, we were able to measure probability densities as small as $10^{-170}$. We also generalize our findings to other stationary Gaussian processes with either diverging, or vanishing  spectral density at zero frequency.
\end{abstract}

\maketitle
\nopagebreak

\section{Introduction}

Large deviations of fluctuating ``dynamical observables", that is of time-integrated quantities, play an important role in non-equilibrium statistical mechanics and probability theory, and they have attracted much attention   \cite{Oono1989,Touchette2009,Hollander,DemboZeitouni,Touchette2018}. Here we focus on a typical class of problems of this type, which deals with large deviations of the scale-invariant time-integrated quantities
\begin{equation}\label{EmpMean}
A_n =\int\limits_{-T}^{T} x(t)^n dt\,,\quad n=1,2,3,\ldots ,
\end{equation}
where $x(t)$ is a centered stationary random process in continuous time.  We will refer to $A_n$ as the area. A closely related quantity is the empirical average of $x^n(t)$, $a_n=A_n/(2T)$. It would be natural to expect that, when $T$ goes to infinity,  the probability density of $A_n$, which we call $\mathcal{P}(A_n,T)$, exhibits a simple large-deviation scaling behavior of the type
\begin{equation}\label{simplescaling}
- \ln \mathcal{P}(A_n,T) \simeq T I(A_n/T)\,,
\end{equation}
where the rate function $I(a_n=A_n/T)$ vanishes at the ensemble-average value of $a_n$. This scaling behavior is considered ``normal" \cite{DemboZeitouni}. It has been shown, however, that even in very simple Markovian systems such as the Ornstein-Uhlenbeck (OU) process \cite{OUpaper}, the ``normal" scaling  (\ref{simplescaling}) holds only for $n=1$ and $2$. In these two cases
the rate function $I(a_n)$  can be obtained, via the Legendre transformation, from the ground state eigenvalue of the tilted generator, entering the Feynman –- Kac equation for the generating function of dynamical average $a_n$, see \textit{e.g.} Ref. \cite{Touchette2018}. Mathematically the problem reduces to that of a quantum harmonic oscillator. For $n>2$ the modified quadratic potential becomes non-confining and has no bound states, signalling a failure of the tilted generator method at sufficiently large $A_n$.
For the large $A_n$, however, the noise is effectively weak. As a result, the probability density $\mathcal{P}(A_n,T)$ is dominated by the optimal path - the most likely trajectory $x(t)$, conditioned on a specified $A_n$ \cite{NT2018}. Finding the optimal path  boils down to minimization of the action functional subject to the constraint  Eq.~(\ref{EmpMean}). This is the essence of the optimal fluctuation method (OFM). Applying it to the cases of $n>2$, Nickelsen and Touchette \cite{NT2018}  showed that  the large-$A_n$ asymptotic of $\mathcal{P}(A_n,T)$ exhibits an ``anomalous" scaling,
\begin{equation}
-\ln \mathcal{P}(a_n\to \infty ,T\to \infty) \simeq c_n (2aT)^{2/n} = c_n A_n^{2/n},
\label{NT3}
\end{equation}
where $c_n=O(1)$.
In the following we will often call $-\ln \mathcal{P}(a_n\to \infty, T) \equiv S(a_n\to \infty, T)$ the action. The anomalous scaling of the action stems from the fact that the optimal path of the conditioned process becomes strongly localized in time, whereas a delocalized solution, which also exists, is suboptimal. This change of the scaling regime due to the optimal paths localization is conceptually similar to the one coming from the ``big-jump principle". The latter is well-known in the context of large deviations of discrete sums of independent identically distributed random variables with a heavy-tailed distribution, see Ref. \cite{Barkai} and references therein.

However, typical fluctuations,  which correspond to not too large $A_n$, 
remain Gaussian even for $n>2$ \cite{Smith2022}. Combining these two facts, Smith \cite{Smith2022} (see also Ref. \cite{Smith2024}) established a complete scaling of $\mathcal{P}(A_n,T)$,
\begin{equation}\label{notsimplescaling}
- \ln \mathcal{P}(A_n,T) \simeq T^{1/(n-1)} I(A_n/T^{n/(2n-2)}),
\end{equation}
where the exponents (the powers of $T$) are both different from 1. Remarkably, this anomalous scaling behavior is accompanied by  a first-order dynamical phase transition -- a finite jump in the first derivative of the function $I(z)$ -- at a critical value $z=z_n=O(1)$ \cite{Smith2022}.

The scaling behavior (\ref{notsimplescaling}) actually extends to a whole class of Gaussian stationary processes. As such processes are in general non-Markovian, the tilted generator technique is unavailable here from the start. Here too the anomalous scaling can be established by combining a Gaussian behavior of typical fluctuations of $A_n$ with an asymptotic of the type of Eq.~(\ref{NT3}) the limit of large $A_n$. The latter follows from (a nonlocal variant of) the OFM which relies on the knowledge of the path integral of Gaussian processes \cite{Meerson2019}.

The nonlocal OFM calculations of Ref. \cite{Meerson2019}, however, relied on the assumptions that the integral over time of the covariance of the Gaussian process, $\int_{-\infty}^{\infty} \nu(\tau) \,d\tau$, is both bounded and nonzero.  In the spectral language this is equivalent to the assumptions that the spectral power density of the process,
$\int_{-\infty}^{\infty} \kappa(\tau) e^{i\omega \tau}\,d\tau$, neither diverges, nor vanishes at $\omega=0$. For brevity, we will refer to  processes obeying these two assumptions as ``ordinary".

Here we extend the previous work by relaxing these assumptions. First, we investigate how long-range correlations of $x(t)$, which cause the divergence of the time integral of the covariance, affect the scaling behavior of large deviations of $A_n$. Second, we study the case when the integral of the covariance vanishes.
A convenient framework for addressing these two questions is the fractional Ornstein-Uhlenbeck  (fOU) process, which has been a subject of active research in its own right, see \textit{e.g.} \cite{Cheredito2003,Kaarakka2015,Metzler2021,Meerson2024}.  The fOU process is a natural extension of the standard OU process, where the white Gaussian noise, driving the system, is replaced by the fractional Gaussian noise (fGn): the time derivative of the fractional Brownian motion with the Hurst exponent $0<H<1$ \cite{Kolmogorov,MvN}. In Section \ref{fOUdescription} we summarize the main properties of the fOU process.

For further comparison,  we recap the known scaling behavior of the action for the ``ordinary" stationary processes:
\begin{equation}
    -\ln \mathcal{P}(A_n=2a_nT\to \infty ,T\to \infty) \sim T^\xi a_n^{2/n},\qquad \xi=\begin{cases}
        1, \qquad & n=1,2\,,\\
        2/n, \qquad & n=3,4,\ldots\\
    \end{cases}
    \label{LDP}
\end{equation}
As we shall see, the scaling behavior (\ref{LDP})  of the $A_n\to \infty$ asymptotic with respect to the integration time $T$ 
breaks down for  both vanishing, and diverging spectral densities at $\omega=0$, that is for $0<H<1/2$ and $1/2<H<1$, respectively.

In Section \ref{DVscaling} we discuss the emergence of an anomalous scaling behavior in the simplest case of
$n=1$. Here the probability distribution of $A_1$ is Gaussian and  can be calculated exactly for any $T$.  We show that the resulting long-time behavior of the action obeys a large-deviation principle of the form
\begin{equation}
    -\ln \mathcal{P}(a_1\to \infty ,T\to \infty )\sim T^{2-2H}a^2_1,\quad 0<H<1\,.
\label{1}
\end{equation}
Here the scaling behavior with $T$ is anomalous except in the special case of $H=1/2$, when the fOU process becomes the standard OU process.

In Section \ref{OFM} we employ the OFM to determine the optimal paths and the large-$A_n$ asymptotics of the action for arbitrary integer $n$. For $n=1$ the OFM calculation reproduces the leading-order behavior~(\ref{1}).

For $n=2$ the action crucially depends on whether the spectral density is bounded or not at zero frequency. If it is bounded (that is, for $H<1/2$), the action follows the simple scaling (\ref{simplescaling}). However, if it diverges (that is, for $H>1/2$), the action exhibits an anomalous scaling in $T$. Overall, we obtain in this case
\begin{equation}\label{2}
  -\ln \mathcal{P}(a_2\to \infty ,T\to \infty )\sim\begin{cases}
       T a_2\,, \qquad & H\leq 1/2;\\
       T^{2-2H} a_2\,, \qquad & H> 1/2.
        \end{cases}
\end{equation}

For $n>2$ the OFM equations have solutions which are localized in time, and delocalized solutions. The competition between them in terms of the action is nontrivial and depends on $H$. As we show here, because of the long-range correlations of the fOU process, the delocalized optimal paths have a smaller action within a certain range of the Hurst exponents. Overall, for $n>2$ we obtain
\begin{equation}\label{3}
     -\ln \mathcal{P}(a_n\to \infty ,T\to \infty )\sim\begin{cases}
        T^{2/n} a_n^{2/n},\qquad & H\leq 1-1/n;\\
        T^{2-2H} a_n^{2/n}, \qquad & H> 1-1/n.
    \end{cases}
\end{equation}

\begin{figure}[ht]
\centering
\includegraphics[clip,width=0.35\textwidth]{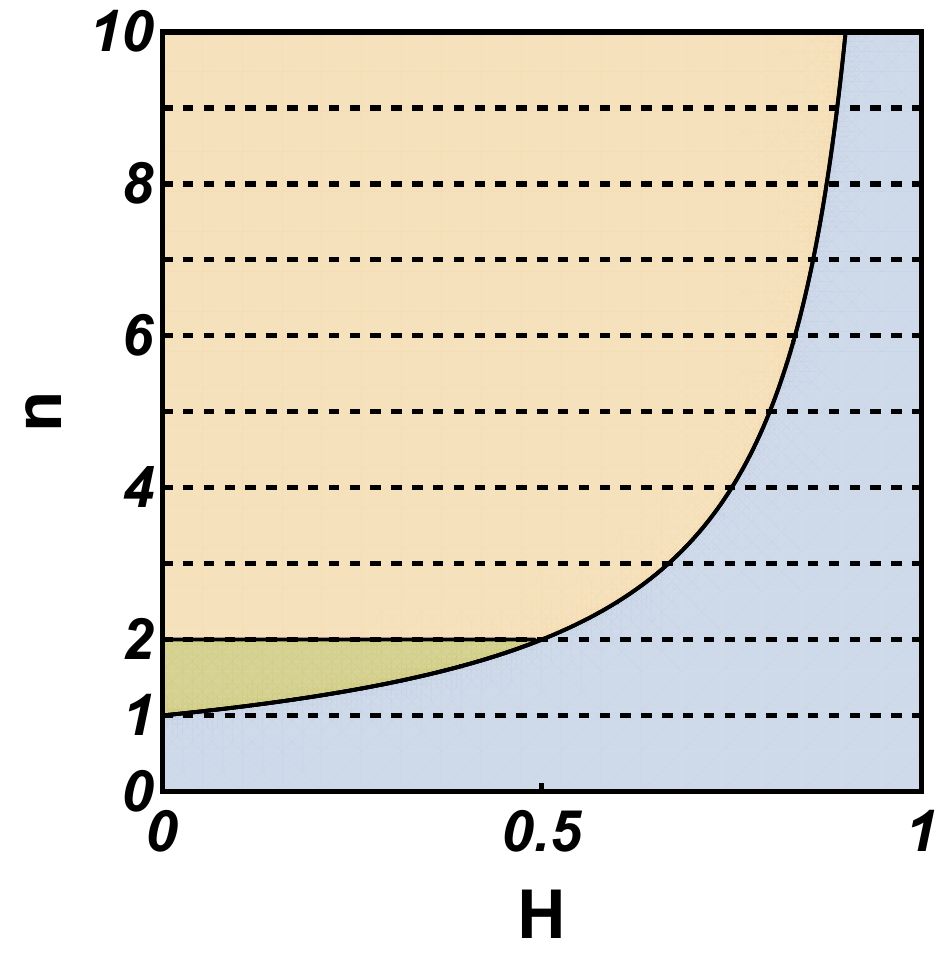}
\caption{Phase diagram of the system on the $(H, n)$ plane. Different colors indicate regions with a different behavior of the exponent $\alpha(H,n)$, determining  the dynamical scaling of the action, $-\ln \mathcal{P}(a_n\to \infty ,T\to \infty )\sim  T^{\alpha(H,n)}\, a^{2/n}_n$, see Eqs.~(\ref{1})-(\ref{3}). The blue region corresponds to a delocalized solution with $\alpha(H,n)=2-2H$. The yellow region corresponds to a strongly localized solution with $\alpha(H,n)=2/n$. The green region corresponds to a delocalized solution with $\alpha(H,n)=1$. For illustrative purposes, we have extended the phase diagram to non-integer $n$. However, our results are applicable only for integer $n\geq 1$, as indicated by the  dashed lines.}
\label{fig_PhaseDiagram}
\end{figure}

The resulting phase diagram on the  $(H, n)$ plane, see Fig.~\ref{fig_PhaseDiagram}, which describes different behaviors of the scaling exponent $\alpha(H,n)$  of the action $-\ln \mathcal{P}(a_n\to \infty ,T\to \infty )\sim  T^{\alpha(H,n)}\, a^{2/n}_n$, is quite nontrivial, as it includes regions dominated by either strongly localized, or delocalized optimal path.

In  Section \ref{Numerics} we verify our main theoretical predictions numerically. To this end we use several methods: a matrix-type discretization for a linear integral equation, an iteration method for solving a nonlinear integral equation, and a multicanonical Monte Carlo sampling method based on the Circulant Embedding Method and the Wang-Landau algorithm. Some technical details are relegated to two Appendices.

\section{Fractional OU process}
\label{fOUdescription}

The fOU process is described by the linear overdamped Langevin equation \cite{Cheredito2003,Kaarakka2015}
\begin{equation}
    \dot{x}(t)+\gamma x=\sqrt{2 D}\xi(t),
    \label{LangevinfOU}
\end{equation}
where $\xi(t)$ if the fGn: a centered stationary Gaussian process with the covariance
\begin{equation}
    c(\tau)=\langle \xi(t+\tau)\xi(t) \rangle=\frac{d}{d\tau}\big(H \vert \tau\vert^{2H-1}\text{sgn}(\tau)\big)=\frac{1}{2}\frac{d^2}{d\tau^2}\vert \tau \vert^{2H}\,.
\label{kappa}
\end{equation}
Here $0<H<1$ is the Hurst exponent. For $H=1/2$, $\xi(t)$ reduces to the white noise: the time derivative of the standard Brownian montion. In this case $c(\tau) =\delta(\tau)$,
and the fOU process becomes a standard OU process.
In the absence of the confining potential, \textit{i.e.} when $\gamma=0$, Eq. (\ref{LangevinfOU}) describes the fractional Brownian motion, where  $H<1/2$ and $H>1/2$ correspond to the subdiffusion and superdiffusion, respectively:
\begin{equation}
    \langle x^2(t)\rangle=2 D |t|^{2H}\,, \quad \text{for $\gamma=0$}.
\end{equation}

The covariance $\kappa(\tau)=\langle x(t+\tau) x(t)\rangle$ of the stationary fOU process can be conveniently calculated by solving Eq.~(\ref{LangevinfOU}) in the Fourier space. The resulting spectral density $\kappa_{\omega}$ of the fOU process is
\begin{equation}
 \kappa_{\omega}=\int\limits_{-\infty}^{\infty}  \kappa(\tau)e^{i \omega \tau}d\tau= \frac{2D c_{\omega}}{\omega^2+\gamma^2}
 =2D \sin(\pi H)\Gamma (2H+1)\frac{\vert \omega \vert^{1-2H}}{\omega^2+\gamma^2}\,,
 \label{kappaomega}
\end{equation}
where
\begin{equation}
 c_{\omega}=\int\limits_{-\infty}^{\infty} c(\tau)e^{i \omega \tau}d\tau=\sin(\pi H)\Gamma (2H+1)\vert \omega \vert^{1-2H}
\label{comega}
\end{equation}
is the spectral density of the fGn $\xi(t)$, and $\Gamma(\ldots)$ is the gamma function.
Applying the inverse Fourier transform to Eq.~(\ref{kappaomega}), one obtains \cite{Cheredito2003}
\begin{equation}
    \kappa(\tau)=\frac{1}{2\pi}\int\limits_{-\infty}^{\infty}
    \kappa_{\omega} e^{-i\omega \tau}d\omega=\frac{D  \Gamma (2 H+1)}{\gamma ^{2 H}} \left[\cosh (\gamma  \tau)-\frac{ |\gamma \tau| ^{2 H} }{\Gamma (2 H+1)}{}_1F_2\left(1;H+\frac{1}{2},H+1;\frac{\tau^2 \gamma ^2}{4}\right)\right],
    \label{cov}
\end{equation}
where $_1F_2(\ldots)$ is the hypergeometric function \cite{Wolfram}. Interestingly, the same result can be obtained by the OFM, see Ref. \cite{Meerson2024} for detail.

In the particular case of $H=1/2$, Eq.~(\ref{cov}) simplifies drastically and gives the familiar expression  for the covariance of the standard OU process:
\begin{equation}\label{covOU}
\kappa(\tau) = \frac{D} {\gamma} e^{- \gamma  |\tau|}.
\end{equation}

As one can see from Eq.~(\ref{comega}), $c_{\omega=0}$ is both bounded and nonzero only for the standard OU process, $H=1/2$. At $H>1/2$, $c_{\omega=0}$ diverges, while at $H<1/2$,  $c_{\omega=0}$ vanishes.  Importantly, these properties are inherited by the spectral density $\kappa_{\omega}$ of the fOU process, see Eq.~(\ref{kappaomega}). As we shall see shortly, these behaviors at zero frequency strongly affect the scaling behavior of the distribution of the dynamical observables: both for the fOU, and for other  stationary Gaussian processes. In particular, we find that the simple scaling~(\ref{simplescaling}) is violated for all $H\neq 1/2$ already for $n=1$.

Applying simple dimensional analysis  \cite{Barenblatt,dimensions}, we can obtain the following exact scaling form of the area distribution:
\begin{equation}
     \mathcal{P}(A_n,T) = \frac{\gamma^{1+Hn}}{D^{n/2}} P\left(\frac{A^{2/n}_n \gamma^{2/n+2H}}{D},\gamma T;H,n\right),\quad \text{or}\quad  \mathcal{P}(a_n,T) =\frac{\gamma^{nH}}{D^{n/2}} P \left(\frac{a^{2/n}_n \gamma^{2H}}{D},\gamma T;H,n\right),
    \label{scaling}
\end{equation}
where the dimensionless function $P(\ldots)$ of its four dimensionless arguments is unknown. We will
calculate the function $P(\ldots)$ exactly only for $n=1$.  For $n>1$ we will employ two large parameters: $T$ and $a_n$. The large parameter $a_n\gg 1$ will enable us to use the OFM, with its characteristic weak-noise scaling form of $\ln \mathcal{P}(A_n,T)$.

\section{The ``normal" scaling breaks down already for $n=1$}
\label{DVscaling}

Since $A_1$ linearly depends on $x(t)$, which is normally distributed, $A_1$ is also normally distributed, and its distribution is fully determined by its mean $\langle A_1\rangle $ (which is zero due to symmetry) and the variance, $\langle A_1^2\rangle$. The latter, for arbitrary integration time $2T$, can be calculated exactly as follows:
\begin{eqnarray}
        &\langle A_1^2\rangle = \int\limits_{-T}^{T}dt\int\limits_{-T}^{T}ds\left\langle x(t)x(s)\right\rangle = \int\limits_{-T}^{T}\int\limits_{-T}^{T}\kappa(t-s)dtds= \nonumber \\
        &= D \Gamma(2H+1)\gamma^{-2H}\left[\frac{4 \sinh ^2(\gamma  T)}{\gamma ^2}-2 \sqrt{\pi } \gamma ^{2 H} T^{2 (H+1)} \, _1\tilde{F}_2\left(1;H+\frac{3}{2},H+2;T^2 \gamma ^2\right)\right]\,,
    \label{W_var}
\end{eqnarray}
where $_1\tilde{F}_2(\ldots)$ is the regularized hypergeometric function, and we evaluated the integral using ``Mathematica" \cite{Wolfram}. Taking the large-$T$ limit, we obtain a simple expression
\begin{eqnarray}
        &\langle A_1^2\rangle \simeq 2 D \gamma ^{-2 (H+1)} (2 T \gamma )^{2 H}=2D \gamma^{-2}(2 T )^{2 H}.
        \label{fOU-mean-scaling}
\end{eqnarray}
As a result, the area  distribution for $n=1$ in the large $T$-limit, and for any $A_1$, has the form
\begin{equation}
    \mathcal{P}(A_1\equiv 2Ta_1,T \gg 1)=\frac{\gamma}{(2 T )^{H}\sqrt{4 \pi D } }\exp\left[-\frac{\gamma^{2} A_1^2}{4 D (2 T )^{2H} }\right]=\frac{\gamma}{(2 T )^{H}\sqrt{4 \pi D } }\exp\left[-(2 T )^{2-2H}\frac{\gamma^{2}a_1^2}{4 D }\right].
    \label{Pmean}
\end{equation}
For $H=1/2$, the expression inside the exponent of Eq.~(\ref{Pmean}) yields the well-known action for the standard OU process [see Eq.~(\ref{NT}) below]. However, the ``normal" scaling of the form (\ref{LDP}) is violated both for $H<1/2$, and for $H>1/2$.

To elucidate the mechanism behind the violation of the ``normal" scaling behavior, let us extend the calculation of the variance~(\ref{W_var}) to an arbitrary stationary Gaussian process $X(t)$ with the covariance $\nu(\tau)=\left\langle X(t+\tau)X(t)\right\rangle$. Going over to the Fourier space, we have
\begin{eqnarray}
    &\langle A_1^2\rangle=\int\limits_{-T}^{T}dt\int\limits_{-T}^{T}ds\left\langle X(t)X(s)\right\rangle=\int\limits_{-T}^{T}dt\int\limits_{-T}^{T}ds\int\limits_{-\infty}^{\infty}\nu_{\omega}e^{-i\omega (t-s)}d\omega=4\int\limits_{-\infty}^\infty \frac{\sin^2(\omega T)}{\omega^2} \nu_\omega d\omega,
    \label{arbnu}
\end{eqnarray}
where $\nu_{\omega}=\int\limits_{-\infty}^{\infty}  \nu(\tau) e^{i \omega \tau}d\tau$ is the spectral density of the process $X(t)$. Differentiating this expression with respect to $T$, we get
\begin{eqnarray}
    &\frac{d\langle A_1^2\rangle}{dT}=4\int\limits_{-\infty}^\infty \frac{\sin(2\omega T)}{\omega} \nu_\omega d\omega.
    \label{int}
\end{eqnarray}
In the large $T$-limit, the function $\sin(2\omega T)/\omega$ rapidly oscillates as a function of $\omega$. As a result, the main contribution to the integral in Eq.~(\ref{int}) is determined by the low-frequency behavior of $\nu_{\omega}$. Suppose that this asymptotic behavior is a power law, $\nu_{\omega\to 0} \sim B|\omega|^\beta$.  Then the large-$T$ asymptotic behavior of the integral~(\ref{int})  is the following:
\begin{eqnarray}
    &\frac{d\langle A_1^2\rangle}{dT}
    \sim B\int\limits_{0}^{1/T} \frac{\sin(2\omega T)}{\omega} \omega^\beta d\omega\sim B T^{-\beta},
    \label{int1}
\end{eqnarray}
where we restrict ourselves to $\beta>-1$. This condition ensures that the total energy of the process is finite, $\int\limits_{-\infty}^\infty \nu_\omega d\omega<\infty$.
Integrating Eq. (\ref{int1}) with respect to $T$, we obtain the  scaling behavior of the area variance
\begin{eqnarray}
    &\langle A_1^2\rangle\sim B T^{1-\beta}.
    \label{int2}
\end{eqnarray}
For $\beta=1-2H$, we reproduce the scaling of $A_1$ of the fOU process, see Eq.~(\ref{fOU-mean-scaling}).
As we can see from Eq.~(\ref{int2}), the ``normal" scaling behavior demands that $\beta=0$, that is the spectral density $\nu_{\omega=0}$ is both bounded and nonzero. Otherwise, an anomalous scaling emerges.

\section{Optimal Fluctuation Method}
\label{OFM}

 \subsection{General}

The starting point of the OFM is the probability density of observing of a given realization $x(t)$. For the fOU  process, which is a stationary Gaussian process, the statistical weight of a realization can be expressed, up to a pre-exponential factor, as  $P[x(t)]\sim \exp(-S[x(t)])$. Here, $S[x(t)]$ is the action functional \cite{ZJ}:
\begin{equation}
        S[x(t)]=\frac{1}{2}\int\limits_{-\infty}^{\infty}dt \int\limits_{-\infty}^{\infty}dt' K(t -t')x(t)x(t'),\label{action_Lan}
\end{equation}
where $K(t)$ is the inverse kernel of the stationary fOU process, defined by the equation
\begin{equation}\label{inversek}
\int\limits_{-\infty}^{\infty}K(\tau)\kappa(\tau-t)d\tau=\delta(t) .
\end{equation}

By conditioning the process  on a large area $A_n$,  the process is ``pushed" into a large-deviation regime, where the action (\ref{action_Lan}) is large. As a result, the large-$A_n$ asymptotic of the action is dominated by the optimal, \textit{i.e.} the most probable, path $x(t)$, which minimizes the action subject to the area constraint  (\ref{EmpMean}). Introducing a Lagrange multiplier $\lambda$ to enforce the area constraint, we proceed to minimize the modified functional \cite{Meerson2019}:
\begin{eqnarray}
     S_{\lambda}[x(t)]=\frac{1}{2}\int\limits_{-\infty}^{\infty}dt \bigg[\int\limits_{-\infty}^{\infty}dt' K(t -t')x(t)x(t')- 2\lambda x^n(t) \big(\theta(t+T)-\theta(t-T)\big)\bigg], \label{Act}
\end{eqnarray}
where $\theta(\dots)$ is the Heaviside step function. The linear variation of the modified action (\ref{Act}) must vanish, which leads to a nonlocal Euler-Lagrange equation for the optimal path $x(t)$ conditioned on a specified area $A_n$:
\begin{equation}
    \int\limits_{-\infty}^{\infty}dt' K(t -t')x(t')=n\lambda x^{n-1}(t)  \big[\theta(t+T)-\theta(t-T)\big].
    \label{IntEq}
\end{equation}
 Multiplying both sides of the equation by the covariance function $\kappa(\tau)$ and using Eq.~(\ref{inversek}), we obtain an equation  in terms of the covariance $\kappa(\dots)$ \cite{Meerson2019}
\begin{equation}
    x(t)=n\lambda \int\limits_{-T}^{T}dt' \kappa(t -t')x^{n-1}(t'),
    \label{IntEqkappa}
\end{equation}
As one can see, there is no need here to explicitly calculate the inverse kernel $K(t-t')$, similarly to some previous applications of the nonlocal OFM \cite{Meerson2019,MeersonOshanin}. On the other hand, due to the non-Markovian nature of the process, the optimal paths, subject to a constraint like Eq.~(\ref{EmpMean}) on a finite interval $|t|<T$, involve both the past $-\infty<t<-T$ and the future $T<t<\infty$ \cite{MeersonOshanin}.

For $n=1$ Eq.~(\ref{IntEqkappa}) provides a ready-to-use formula for the calculation of the (unique) optimal path $x(t)$. However, for $n=2$, Eq.~(\ref{IntEqkappa}) is a linear integral equation,  which can be interpreted as an eigenvalue problem and has an infinite number of eigenvalues and corresponding eigenfunctions.   For $n>2$ the integral equation~(\ref{IntEqkappa}) is nonlinear, and there can be multiple solutions \cite{Meerson2019}. In  such cases the least-action solution should be selected.

For $|t|\gg T$,  where $T$ is finite -- that is, well beyond the integration interval $|t|<T$ -- the optimal path is determined by the large-$z$ asymptotics of the covariance $\kappa(z)$:
\begin{equation}
 x(t\gg T)=n\lambda \int\limits_{-T}^{T}dt' \kappa(t -t')x^{n-1}(t')\simeq n\lambda \kappa(t) \int\limits_{-T}^{T}dt' x^{n-1}(t')\sim \kappa(t)\,.
 \label{xlarget}
\end{equation}
Analyzing Eq.~(\ref{cov}), one can see that for $H\neq 1/2$ the decay proceeds a power law: $x(t\gg T)\sim  |t|^{2H-2}$. For $H=1/2$ $x(t)$ decays as $\exp(-\gamma |t|)$.

Using Eqs.~(\ref{action_Lan}) and (\ref{IntEq}), we obtain the action $S(A_n,T)$:
\begin{eqnarray}
    S(A_n,T)=\frac{1}{2}\int\limits_{-\infty}^{\infty}dt x(t)\int\limits_{-\infty}^{\infty}dt' K(t -t')x(t')=\frac{n\lambda}{2}\int\limits_{-T}^{T}dt x^n(t)=\frac{n\lambda A_n}{2},
    \label{act-labmda}
\end{eqnarray}
where the Lagrange multiplier $\lambda$ is to be determined from Eq.~(\ref{EmpMean}) \cite{Meerson2019}.

The inverse kernel $K(t)$ and, as a result,  the OFM action (\ref{action_Lan}) scale as $D^{-1}$. As a result,
we can push the dimensional analysis of Section \ref{fOUdescription} further and argue that the action  has the following scaling behavior:
\begin{equation}
     S(A_n,T) =\frac{\gamma^{2/n+2H} A^{2/n}_n }{D}\Phi\left(\gamma T;H,n\right),\quad \text{or}\quad  S(a_n,T) =\frac{\gamma^{2H}a^{2/n}_n}{D}\Phi\left(\gamma T;H,n\right),
    \label{scaling-action}
\end{equation}
where $\Phi(\ldots)$ is a dimensionless function of its three dimensionless  arguments. We notice immediately that the scaling $S \sim a_n^{2/n}$ is universal, that is $H$-independent.

\subsection{Back to $H=1/2$}

To provide some important reference points for the remainder of the paper, let us briefly review the known results for $H=1/2$. In this case, as we briefly mentioned in the Introduction, the full rate functions for $n=1$ and $2$ can be derived from the dominant eigenvalue of the Feynman-Kac equation for the generating function of the area $A_n$. To probe the cases of $n>2$, where the ``tilted operator" approach fails,  Nickelsen and Touchette \cite{NT2018} employed the OFM which applies for large $a_n$ when the noise is effectively weak. Since  for $H=1/2$ the process is local in time, the minimization of the action $\frac{1}{4D}\int [\dot{x}(t)+\gamma x(t)]^2dt$ subject to the area constraint ~(\ref{EmpMean}) brings about an Euler-Lagrange equation of effective  classical mechanics:
\begin{equation}
    \ddot{x}(t)-\gamma^2 x(t)=-2D n\lambda x^{n-1}(t)\big[\theta(t+T)-\theta(t-T)\big]\,.
\label{h12}
\end{equation}
Solving this equation for $n=1$ and $n=2$ and taking the limit of $T\to\infty$, one can see that the optimal path stays close to the fixed point [$x(t)=A_1/2T=a_1$ and $x(t)=\pm\sqrt{A_2/2T}=\pm\sqrt{a_2}$, for $n=1$ and $2$, respectively] on the phase plane $(x,\dot{x})$ \cite{NT2018}. These almost constant-in-time solutions determine the large-$a_n$ asymptotic of the action:
\begin{equation}
    -\ln \mathcal{P}(a_n\to \infty ,T\to \infty) \simeq 2T I(a_n),\qquad I(a_n)=\begin{cases}
        \frac{ \gamma^2 a_1^2}{4 D}, \qquad & \text{for $n=1$,}\\
        \frac{ \gamma^2 a_2}{4 D}, \qquad & \text{for $n=2$.}\\
    \end{cases}
    \label{NT}
\end{equation}
A constant solution to Eq. (\ref{h12}) at $T\to \infty$ exists for any $n$. However, for $n>2$ a different solution appears, which has a lesser action \cite{NT2018}. At $T\to \infty$ this solution corresponds to a homoclinic orbit on the phase plane $(x,\dot{x})$. This solution is referred to as an instanton, and its explicit analytic form was found in Ref. \cite{Meerson2019}:
\begin{equation}
    x(t)=\left(\frac{4 D \lambda }{\gamma ^2}\right)^{\frac{1}{2-n}} \text{sech}^{\frac{2}{n-2}}\left[\frac{  (n-2) \gamma t}{2} \right] .
    \label{OUx(t)}
\end{equation}
The instanton is strongly localized on the time scale of the correlation time $1/\gamma$. Determining $\lambda$ and calculating the action, one arrives at Eq.~(\ref{NT3}) \cite{NT2018}.

\subsection{$n=1$}
\label{OFM1}

The case $n=1$, where we have exact results from Section \ref{DVscaling}, provides a useful check of the OFM.
As $A_1$ is normally distributed, we can restore the complete long-time probability distribution $\mathcal{P}(A_1,T)$ from the knowledge of its large-$A_1$ tail. The optimal path $x(t)$, dominating this tail, is described by Eq. (\ref{IntEqkappa}) with $n=1$:
\begin{equation}
    x(t)=\lambda \int\limits_{-T}^{T}dt' \kappa(t -t').
\label{integral1}
\end{equation}
It is convenient to evaluate the  integral in Eq.~(\ref{IntEqkappa}) by Fourier-transforming the equation. In this way we obtain
\begin{equation}
    x(\omega)= \frac{2 \lambda   \kappa_{\omega} \sin (\omega T)}{\omega},
    \label{n=1FT}
\end{equation}
where  the spectral density of the fOU $\kappa_{\omega}$ is given by Eq.~(\ref{kappaomega}).
In the large-$T$ limit, the function $\sin(\omega T)/\omega$ in Eq. (\ref{n=1FT}) rapidly oscillates as a function of $\omega$. Therefore the inverse Fourier transform of $x(\omega)$ is dominated by the low-frequency behavior of the spectral density (\ref{kappaomega}):
\begin{equation}
    \kappa_{\omega\to 0}\simeq \frac{2D}{\gamma^2} \sin(\pi H)\Gamma (2H+1) \vert \omega \vert^{1-2H}.
\end{equation}
Plugging this asymptotic into Eq. (\ref{n=1FT}) and performing the inverse Fourier transform, we obtain the large-$T$ asymptotic of the optimal path:
\begin{eqnarray}
     &x(t)=\frac{2 D \lambda H }{\gamma^2}\left[\text{sgn}(t+T) | t+T| ^{2 H-1}-\text{sgn}(t-T) | t-T| ^{2 H-1}\right],\quad H\neq 1/2.
     \label{fOU-x}
\end{eqnarray}
Now Eq.~(\ref{EmpMean}) yields the Lagrange multiplier $\lambda$ in terms of $A_1$:
\begin{equation}
\lambda= \frac{\gamma^{2}A_1}{2 D  (2 T )^{2H}},
\label{n=1Lambda}
\end{equation}
and we obtain the optimal path for large $T$:
\begin{equation}
    x(t)=\frac{A_1 H}{ (2T)^{2 H}} \left[\text{sgn}(t+T) | t+T| ^{2 H-1}-\text{sgn}(t-T) | t-T| ^{2 H-1}\right],\quad H\neq 1/2.
    \label{optn=1}
\end{equation}
For $H=1/2$ we can evaluate the integral in Eq.~(\ref{integral1}) exactly by using Eq.~(\ref{covOU}). We obtain
 \begin{equation}
 x(t)=\begin{cases}
      \frac{A_1 \gamma  \left(1-e^{-\gamma T} \cosh (\gamma  t)\right)}{2 \gamma  T+e^{-2 \gamma  T}-1}, \qquad & \text{for $|t|\leq T$,}\\
      \frac{A_1 \gamma  e^{-\gamma  |t|} \sinh (\gamma  T)}{2 \gamma  T+e^{-2 \gamma  T}-1}, \qquad & \text{for $|t|> T$.}
 \end{cases}
 \end{equation}
For $\gamma T\gg 1$ this optimal path stays very close to the constant value $a_1=A_1/2T$ within the interval $|t|<T$ (in agreement with Ref.~\cite{NT2018}), and it decays exponentially fast outside the interval. Actually, Eq.~(\ref{optn=1}) approximately describes the optimal path for $H=1/2$ as well beyond the narrow boundary layers of width $O(1/\gamma)\ll T$ around $|t|=T$.

The $|t|\gg T$ asymptotic of the optimal path (\ref{optn=1}),
\begin{equation}
    x(|t|\gg T)\simeq \frac{A_1 H (2H-1)}{2^{2H}T}\left|\frac{t}{T}\right|^{2H-2},\quad H\neq 1/2.
\end{equation}
is governed by the large-$\tau$ asymptotics of $\kappa(\tau)$, as to be expected from Eq.~(\ref{xlarget}).

\begin{figure}[ht]
\centering
\includegraphics[clip,width=0.48\textwidth]{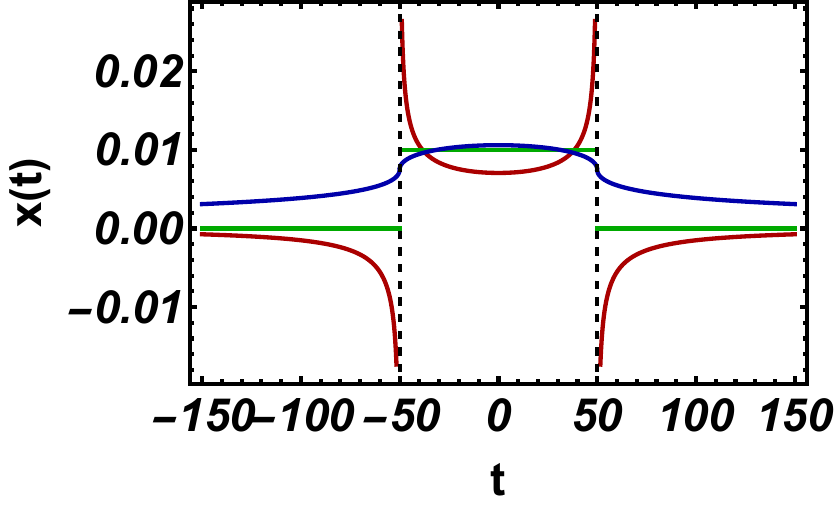}
\caption{Optimal paths $x(t)$ for $n=1$, as described by Eq.~(\ref{optn=1}), for $H=1/4$ (red), $H=1/2$ (green), and $H=3/4$ (blue) for $T=50$.}
\label{fig1}
\end{figure}

Some examples of the optimal paths (\ref{optn=1})  for different values of $H$ are shown in Fig.~\ref{fig1}. As one can see, the optimal paths provide a valuable insight into the mechanism of large deviations  of the area at different $H$. For $H=1/2$, the optimal path on the interval $|t|<T$ stays close to the fixed point of the local classical mechanics. For $H<1/2$, the optimal path exhibits an anti-persistent behavior, with pronounced rise or fall near $t=\pm T$. This behavior reflects the anti-correlated nature of the fOU process for $H<1/2$. In contrast, for $H>1/2$, the optimal path exhibits persistence by retaining its direction over longer periods of time, which results in smoother, more monotonic behavior. This reflects the long-range correlations of the fOU process for $H>1/2$.

Substituting the Largange multiplier (\ref{n=1Lambda}) into  Eq. (\ref{act-labmda}),  we arrive at the large-$T$ asymptotic of the action:
\begin{equation}
    S(A_1,T)=\frac{\gamma^{2}A_1^2}{4 D (2 T)^{2H} }=(2 T )^{2-2H}\frac{\gamma^{2}a_1^2}{4 D }.
    \label{Sn=1}
\end{equation}
As to be expected, this action coincides with the expression inside the exponent of the exact distribution  (\ref{Pmean}).

\subsection{$n=2$}
\label{OFM2}
For $n = 2$, Eq.~(\ref{IntEqkappa}) for the optimal path $x(t)$ is a homogeneous linear Fredholm equation of the second kind:
\begin{equation}
    x(t)=2\lambda \int\limits_{-T}^{T}dt' \kappa(t -t')x(t'),
    \label{n=2FE}
\end{equation}
Equation~(\ref{n=2FE}) can be viewed as an eigenvalue problem, the Lagrange parameter $\lambda$ playing the role of the eigenvalue \cite{Meerson2019}. There is an infinite number of the eigenvalues and eigenfunctions.
Here too, as we will demonstrate shortly, the distinction between short and long correlations is crucial.

\subsubsection{$H\leq \frac{1}{2}$}

\begin{figure}[ht]
\centering
\includegraphics[clip,width=0.3\textwidth]{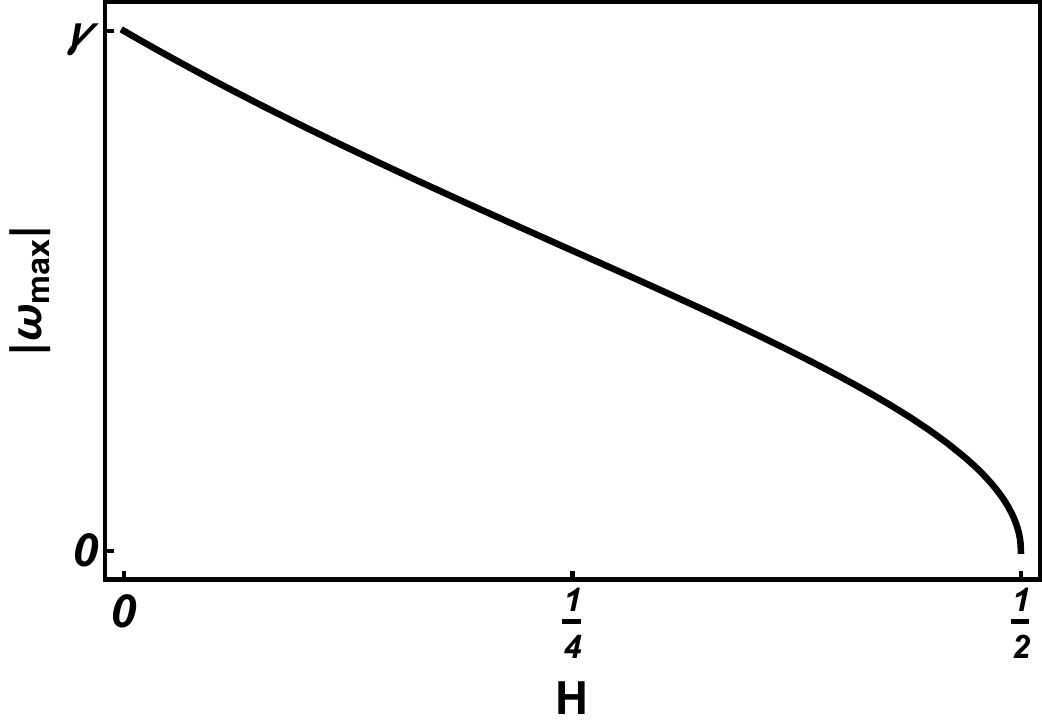}
\caption{Location of the maximum $|\omega_{\text{max}}|=\gamma  \sqrt{\frac{1-2H}{1+2 H}}$ of the covariance's spectral density (\ref{cov}).}
\label{fig_max}
\end{figure}

For short correlations, that is for covariances with a spectral density bounded at $\omega=0$, the limits of integration at large $T$ in Eq.~(\ref{n=2FE}) can be sent to infinity \cite{Meerson2019}. Hence, for $H\leq 1/2$, Eq.~(\ref{n=2FE}) becomes
\begin{equation}
    x(t)=2\lambda \int\limits_{-\infty}^{\infty}dt' \kappa(t -t')x(t'),
    \label{n=2}
\end{equation}
which in the Fourier space simplifies to
\begin{equation}
   2\lambda \kappa_\omega =1.
    \label{n=2FT}
\end{equation}
This yields a continuum spectrum of eigenvalues $\lambda=(2 \kappa_\omega)^{-1}$. For each eigenvalue there are two associated linearly independent eigenfunctions: $x(t)=\{\cos(\omega t), \sin(\omega t)\}$. The general solution of Eq.~(\ref{n=2}) is a linear combination of the eigenfunctions. We need, however, to minimize the action (\ref{act-labmda}).  For that we must select the minimal eigenvalue. The latter is determined by the maximum of the spectral density (\ref{cov}). As one can check, the maxima are located at
\begin{equation}\label{omegamax}
\omega_{\text{max}}=\pm\gamma  \sqrt{\frac{1-2H}{1+2 H}} ,
\end{equation}
see Fig.~\ref{fig_max}. Substituting these into Eq. (\ref{n=2FT}), we obtain
\begin{equation}
    \lambda=\frac{ \gamma ^{2 H+1} \csc (\pi  H)}{2 D \sqrt{1-4 H^2} \Gamma (2 H+1)}\left(\frac{1-2H}{1+2 H}\right)^H,\quad 0<H\leq 1/2.
    \label{lambda_min}
\end{equation}
Then, combining Eqs.~(\ref{act-labmda}) and~(\ref{lambda_min}), we arrive at the action for $H\leq 1/2$:
\begin{equation}
    S(A_2)=\frac{ \gamma ^{2 H+1} \csc (\pi  H)}{2 D \sqrt{1-4 H^2} \,\Gamma (2 H+1)} \left(\frac{1-2H}{1+2 H}\right)^H A_2.
    \label{Sbound}
\end{equation}

In the particular case of $H=1/2$, Eq.~(\ref{Sbound}) simplifies to $S(A_2)=\frac{\gamma^2A_2}{4 D}$ and perfectly agrees with the large-$A_2$ asymptotics of the action of the quadratic observable of the OU process (\ref{NT}). Besides, for $H=1/2$, $\kappa_{\omega}$ has a single maximum at $\omega=0$. As a result, in the $T\to\infty$ limit, the corresponding eigenfunction $x(t)$ tends to a constant, as to be expected.

\subsubsection{$H>\frac{1}{2}$}
\label{n2H12}

For $H > 1/2$, $\kappa_{\omega}$ has a single maximum, in the form of an integrable singularity, at $\omega=0$. As a result, the relation $\lambda=1/(2\kappa_{\omega})$ would only give a trivial solution. This signals that one is not allowed to extend the integration limits in Eq. (\ref{n=2FE}) to infinity. For finite $T$, the minimal eigenvalue obeys the relation \cite{Polyanin2008}
\begin{equation}
    \frac{1}{2\lambda}=\underset{x\neq 0}{\text{max}}\frac{\int\limits_{-T}^{T}\int\limits_{-T}^{T}dt' dt \kappa(t -t')x(t')x(t)}{\int\limits_{-T}^{T}dt x^2(t)},
    \label{n=2min}
\end{equation}
and the maximum of the r.h.s. is achieved when $x(t)$ is the corresponding eigenfunction. For  $H=1/2$, where finite-$T$ corrections can be easily evaluated analytically (see Appendix \ref{appA}), we find that the leading-order contribution to the action is dominated by constant solutions for the optimal paths. Extending this argument to the $H>1/2$ case, we will assume that, in large $T$ limit, the eigenfunction $x(t)$, corresponding to the minimal eigenvalue, also become nearly flat, and therefore can be accurately approximated by a constant solution $x(|t| < T) = \pm\sqrt{A_2/2T}$ which fulfills the area constraint (\ref{EmpMean}). We will verify this assumption numerically
in Section~\ref{Numerics}. Substituting this constant into Eq. (\ref{n=2min}), we obtain
\begin{equation}
    \frac{1}{\lambda}\simeq \frac{\int\limits_{-T}^{T}\int\limits_{-T}^{T}\kappa(t-s)dtds}{T}.
\end{equation}
Using large $T$ asymptotics of the integral in the r.h.s., see Eqs.~(\ref{W_var}) and~(\ref{fOU-mean-scaling}),  we determine the Lagrange multiplier:
\begin{equation}
    \lambda\simeq T\left(\int\limits_{-T}^{T}\int\limits_{-T}^{T}\kappa(t-s)dtds\right)^{-1}\simeq \frac{\gamma^{2} T^{1-2H}}{2^{2H+1}D}.
    \label{lamH}
\end{equation}
Combining Eqs.~(\ref{act-labmda}) and~(\ref{lamH}), we arrive at the action
\begin{equation}
    S(A_2,T)=\frac{\gamma^{2} T^{1-2H} A_2}{2^{2H+1}D }=\frac{\gamma^{2}T^{2-2H}a_2}{2^{2H}D},\qquad H\geq 1/2.
    \label{n=2_act}
\end{equation}
For $H = 1/2$ Eq.~(\ref{n=2_act}) coincides with  Eqs. (\ref{Sbound}) and  (\ref{NT}).

Combining Eqs. (\ref{Sbound}) and (\ref{n=2_act}), we obtain
\begin{equation}
    S(A_2,T)\sim\begin{cases}
        A_2,\qquad & H\leq 1/2;\\
       T^{1-2H} A_2 , \qquad & H> 1/2,
    \end{cases} \quad \text{or} \quad S(a_2,T)\sim\begin{cases}
        T a_2  ,\qquad & H\leq 1/2;\\
        T^{2-2H} a_2 , \qquad & H> 1/2.
    \end{cases}
    \label{A_T_n=2}
\end{equation}
As we can see, if the spectral density is finite ($H\leq 1/2$), the action follows the simple scaling (\ref{simplescaling}). However, if the density is singular ($H>1/2$), the action has an anomalous scaling in $T$.
In both cases, the optimal paths are determined by the position of the spectral density maximum, and they are delocalized. For $H>1/2$ they  can be approximated by the relevant constant, while for $H\leq 1/2$ they are oscillating and given by a linear combination of the harmonic functions $\cos(\omega_{\text{max}} t)$ and $\sin(\omega_{\text{max}} t)$, where $\omega_{\text{max}}$ is described by Eq.~(\ref{omegamax}).

\subsubsection{A generalization}
\label{general}

Based on these results, we can argue that, for an arbitrary stationary Gaussian process $X(t)$  with covariance $\nu(t)$, whose spectral density  is singular at zero frequency,  $\nu_{\omega\to 0}\simeq B|\omega|^\beta$, where $-1<\beta<0$,   the action can be determined by using the constant approximation of the optimal path $x(|t| < T) = \pm\sqrt{A_2/2T}$. As a result, the corresponding Lagrange multiplier can be estimated using Eq. (\ref{n=2min}):
\begin{equation}
    \lambda = T \left( \int\limits_{-T}^{T} \int\limits_{-T}^{T} \nu(t - s) \, dt \, ds \right)^{-1}.
    \label{arb_lam}
\end{equation}
The  double integral in the r.h.s. of Eq. (\ref{arb_lam}) is again dominated by the low-frequency behavior of the noise spectral density. Repeating  the calculations in Eqs. (\ref{arbnu}) –- (\ref{int2}) gives the scaling behavior of the Lagrange multiplier for large $T$:
\[
\lambda  \simeq B^{-1} T^{\beta}.
\label{int_alpha}
\]
Alongside with Eq. (\ref{act-labmda}), this gives an anomalous scaling behavior of the action at large $T$:
\begin{eqnarray}
    S(A_2,T)=\lambda A_2\sim B^{-1}T^{\beta} A_2\sim B^{-1}T^{1+\beta }a_2,
\end{eqnarray}
which, for $\beta=1-2H$, matches the scaling behavior of the action given in Eq. (\ref{n=2_act}).

\subsection{$n>2$}
\label{SectionHM}

For $n>2 $ the integral eqution~(\ref{IntEqkappa}) for the optimal path is nonlinear, and
it can have multiple solutions with different actions. The dimensionless change of  variable $y(t)=(n\lambda)^\frac{1}{n-2}x(t)$  transforms Eq.~(\ref{IntEqkappa}) into a $\lambda$-independent form \cite{Meerson2019},
\begin{equation}
    y(t)=\int\limits_{-T}^{T}dt' \kappa(t -t')y^{n-1}(t'),
    \label{yn}
\end{equation}
while the area constraint (\ref{EmpMean}) becomes
\begin{equation}
    A_n=(n\lambda)^{-\frac{n}{n-2}}\int\limits_{-T}^{T}dt\, y^{n}(t).
    \label{constraint}
\end{equation}
Although for some particular covariances exact analytical solutions of this nonlinear problem are possible \cite{Meerson2019}, the case of fOU  does not seem to be analytically solvable. In Section \ref{Numerics} we solve this problem for several values of $n>2$ numerically, by iterations. We also perform the Wang--Landau Monte Carlo simulations of the original stochastic problem.  But before we proceed to the numerics, let us analyze some trial functions for $x(t)$ and try to determine the scaling behavior of the action.

As we know,  for processes with finite spectral densities, the optimal paths  for large $T$ and for $n>2$ are strongly localized on a $T$-independent time scale,  while (almost) constant solutions still exist but are suboptimal
\cite{NT2018,Meerson2019}. In this subsection we will explore some possible scenarios associated with localized and nonlocalized trial paths for the fOU process. These will also include the possibility of an ``intermediate localization" of the path on a time scale which does grow with $T$, but slower than linearly.

First let us consider a trial path, localized on a $T$-independent time scale $O(1/\gamma)$. In this case, the area scales as
\begin{equation}
    A_n \sim \lambda^{-\frac{n}{n-2}}\int\limits_{-\tau}^{\tau} y^{n}( t) dt,
\end{equation}
and is independent of $T$. The corresponding $T$-independent Lagrange multiplier is
\begin{equation}
\lambda\sim\left(A_n^{-1}\int\limits_{-\tau}^{\tau} y^{n}( t) dt  \right)^{\frac{n-2}{n}}\,.
\end{equation}
This results  in the following  scaling behavior of the action:
\begin{equation}
    S(A_n,T)=f(H,n)\frac{A_n^{2/n}\gamma^{2H+2/n}}{D},
    \label{act-local}
\end{equation}
where $f(H,n)$ is a dimensionless function. Equation~(\ref{act-local}) is manifestly independent of $T$.

Delocalized and intermediately localized trial paths  extend over a time scale which grows with $T$. They can be expected to occur only for $H>1/2$, where the covariance $\kappa(t)$ has a heavy tail, which is scale invariant: 
\begin{equation}
\kappa(t \to \infty) \simeq \kappa_\text{as}(t)=\frac{2D H(2H-1)}{\gamma^2} |t|^{2H-2}.
\label{kas}
\end{equation}
This scale invariance should reflect itself in the optimal path $y(t)$. Therefore, we assume a self-similar behavior of $y(t)$ with $T$,
\begin{equation}
y(t) = T^{\zeta} u(t/T^\chi),
\label{ans1}
\end{equation}
where the exponents $0<\chi\leq 1$ and $\zeta$ are \textit{a priori} unknown. We also assume that the dominant contribution to the integral in Eq.~(\ref{yn}) comes from the timescale of order $\mathcal{O}(T^\chi)$. Plugging the ansatz~(\ref{ans1})  into Eq.~(\ref{yn}) and introducing rescaled variables $w=\tau/T^\chi$ and $z=t/T^\chi$, we obtain
\begin{equation}
    T^\zeta u(z)=\frac{2D H(2H-1)}{\gamma^2} T^{\zeta(n-1)+\chi(2H-1)}\int\limits_{-1}^{1} u^{n-1}(w) |z-w|^{2H-2}dw.
    \label{u_scaling}
\end{equation}
Assuming that the integral on the r.h.s. converges and getting rid of the powers of $T$, we find the following relation between the exponents $\zeta$ and $\chi$:
\begin{eqnarray}
    \zeta=\chi\frac{1-2H}{n-2}.
    \label{ans2}
\end{eqnarray}
Note that, for $n>2$ and $H>1/2$, $\zeta$ is negative. Now we plug the ansatz (\ref{ans1}) with this $\zeta$ into Eq.~(\ref{EmpMean}):
\begin{equation}
    A_n=\int\limits_{-T}^{T} x^{n}(t) dt=(n\lambda)^{-\frac{n}{n-2}}\int\limits_{-T}^{T} y^{n}(t) dt =(n\lambda)^{-\frac{n}{n-2}}T^{\chi\frac{2n(1-H)-2}{n-2}}\int\limits_{-1}^{1} u^{n}(z) dz.
\end{equation}
Expressing the corresponding Lagrange multiplier $\lambda$ through $A_n$ and $T$, 
and  substituting it into Eq. (\ref{act-labmda}), we obtain the scaling behavior of the action for the delocalized and intermediately localized paths:
\begin{equation}
    S(A_n,T)=\frac{1}{2}T^{2\chi(1-H-\frac{1}{n})}A_n^{\frac{2}{n}}\left(\,\int\limits_{-1}^{1} u^n(z)dz\right)^{\frac{n-2}{n}}=g(H,n) T^{2\chi(1-H-\frac{1}{n})}  \frac{\gamma^{2\chi+2(1-\chi)\left(H+\frac{1}{n}\right)} A_n^{\frac{2}{n}}}{D},
    \label{action_slow}
\end{equation}
where $g(H,n)$ is another dimensionless function.  The optimal path, in the original variables, has the following scaling behavior:
\begin{equation}\label{inter_x}
    x(t)\sim A^{\frac{1}{n}}_n T^{-\frac{\chi}{n}} u\left(\frac{t}{T^\chi}\right).
\end{equation}

As one can see from Eq.~(\ref{action_slow}),  the $A_n^{2/n}$ dependence of the action remains the same as in the simple scaling~(\ref{LDP}), as to be expected from the dimensional analysis, see Eq.~(\ref{scaling-action}). Another salient feature of Eq.~(\ref{action_slow}) is that the action grows with $T$ for $H<1-1/n$, is constant at $H=1-1/n$, and decreases for $H>1-1/n$. These behaviors hold for any exponent $0<\chi<1$, but the exponent itself remains undetermined at this stage. This raises an interesting question about the selection rule for the true value of the exponent.
As we will demonstrate numerically in Section \ref{Numerics}, for $H<1-1/n$,  the action (\ref{action_slow}) is $T$-independent, implying that $\chi=0$, and the optimal path is fully localized. For $H>1-1/n$ we find that $\chi=1$: the optimal path is fully delocalized over the entire integration interval. Remarkably, for the specific values of $\chi=0$ and $\chi=1$,  the  associated scaling behavior of the action with $T$, $S(A_n,T)\sim A^{2/n}_n$ for $H<1-1/n$ and $S(A_n,T)\sim T^{2-2H-2/n} A^{2/n}_n $ for $H>1-1/n$, are the slowest possible. Essentially,  both $\chi=0$ and $\chi=1$ are selected by the minimum action. Only at the borderline value $H=1-1/n$, where $\chi$ does not affect the scaling behavior of the action, our numerics show an intermediate localization with $\chi$ apparently equal to $1/3$. In the rest of this subsection we will explore the fully delocalized paths with $\chi=1$.

An additional insight into the delocalized trial path scenario with $\chi=1$  is provided by the direct path-integral representation for the Langevin equation~(\ref{LangevinfOU}) which describes the fOU process. The corresponding action functional is the following:
\begin{eqnarray}
     S[x(t)]=\frac{1}{4D}\int\limits_{-\infty}^{\infty}dt \int\limits_{-\infty}^{\infty}dt' C(t -t')\big[\dot{x}(t)+\gamma x(t)\big]\big[\dot{x}(t')+\gamma x(t')\big],
     \label{Act-xi}
\end{eqnarray}
where $C(t)$ is the inverse kernel, defined by the relation $\int_{-\infty}^{\infty}C(\tau)c(\tau-t)d\tau=\delta(t)$, and $c(\tau)$ is the covariance of the fGn, see Eq.~(\ref{kappa}). The kernel $C(t)$ is known explicitly \cite{MBO}.

Introducing a Lagrange multiplier $\lambda$ as in Eq.~(\ref{Act}) and minimizing the modified action functional, one arrives at an integro-differential equation, which describes the same constrained optimal path $x(t)$ that we have dealt with above, but in terms of the fGn:
\begin{equation}
 \ddot{x}(t) -\gamma^2 x(t)=-2D n\lambda   \int\limits_{-T}^{T}dt' c(t -t')x^{n-1}(t').
 \label{ideq}
\end{equation}

Using the same dimensionless change of  variable $y(t)=(n\lambda)^\frac{1}{n-2}x(t)$ as before, we can rewrite Eq.~(\ref{ideq}) as
\begin{equation}
\ddot{y}(t)-\gamma^2 y(t)=-2D\int\limits_{-T}^{T}dt' c(t -t') y^{n-1}(t').
\label{nonst}
\end{equation}
By applying  the ansatz~(\ref{ans1}) with $\chi=1$ for a delocalized optimal path, we see that, at large $T$, the ``acceleration" term, which scales as $\ddot{y}(t)\sim \mathcal{O}(T^{\zeta-2})$ can be neglected. [To remind the reader, $\zeta<0$, see Eq.~(\ref{ans2})]. Substituting the ansatz (\ref{ans1}) and the covariance of the fGn, $c(t)=H(2H-1)|t|^{2H-2}$, into Eq. (\ref{nonst}), we recover Eq.~(\ref{u_scaling}). Therefore, we can argue that a small correction which we neglected by using the ansatz~(\ref{ans1}), is of order $\mathcal{O}(T^{-2})$.

Further, in the delocalized scenario, the ``velocity" $\dot{y}(t)=\mathcal{O}(T^{\zeta-1})$, and it becomes negligible at large $T$. Hence, we can assume that our slowly varying function $u(t/T)$ is almost a constant: $u(t/T)\simeq \text{const}$. Substituting $u=\text{const}$ into Eq. (\ref{u_scaling}), eliminating the $T$-dependence and integrating the both part of the equation over the interval $(-1,1)$, we obtain
\begin{equation}
   u^{n-2}(t/T)\simeq (\text{const})^{n-2}=\left(\frac{D H(2H-1)}{\gamma^2}\int\limits_{-1}^{1}\int\limits_{-1}^{1} |z-w|^{2H-2}dw dz\right)^{-1}=\frac{\gamma^{2}}{2^{2H} D},
\end{equation}
Substituting this constant solution into Eq. (\ref{action_slow}), we get the following estimate of the action of a delocalized solution with $\chi=1$:
\begin{equation}
    S(A_n,T)=T^{2-2H-\frac{2}{n}}\frac{\gamma^{2} A_n^{\frac{2}{n}}}{2^{2H+\frac{2}{n}} D}.
    \label{action_const}
\end{equation}

Combining Eqs. (\ref{act-local}), (\ref{action_slow})  and (\ref{action_const}),  we obtain the predicted scaling behavior of the action for $n>2$:
\begin{equation}
    S(A_n,T)\sim\begin{cases}
        A_n^{2/n},\qquad & 0<H< \frac{n-1}{n};\\
        T^{2-2H-2/n} A_n^{2/n}, \qquad & H\geq \frac{n-1}{n},
    \end{cases} \quad \text{or} \quad S(a_n,T)\sim\begin{cases}
        T^{2/n}a_n^{2/n} ,\qquad & 0<H<\frac{n-1}{n};\\
        T^{2-2H}a_n^{2/n}, \qquad & H\geq \frac{n-1}{n}\,,
    \end{cases}
    \label{A_T}
\end{equation}
where we classify the case $H=1-1/n$, where the action is independent of $T$, as a delocalized solution. Note that the expressions in first line of Eq.~(\ref{A_T}) are valid down to $H=0$.

Similar arguments can be applied to an arbitrary stationary Gaussian process   $X(t)$ which has a power-law singularity of the spectral density at zero frequency: $\nu_{\omega\to 0}\simeq B|\omega|^\beta$, where $-1<\beta<0$. Assuming the presence of a $T$-independent intrinsic time scale in the problem (an analog of $1/\gamma$),  we can predict the critical value of $n$ at which the transition from a delocalized to a localized solution occurs. Here we assume that, in the large-$T$ limit, there is a  delocalized solution to Eq. (\ref{yn})  with a covariance $\nu(t)$ and $\chi=1$, and it is well-approximated by a constant. The latter is again determined by integrating both part of Eq. (\ref{yn}) over the interval $(-T,T)$:
\begin{equation}
    (\text{const})^{(n-2)}=2T \left( \int\limits_{-T}^{T} \int\limits_{-T}^{T} \nu(t - s) \, dt \, ds \right)^{-1} \sim B^{-1} T^{\beta},
\end{equation}
where we again assumed that the integral is dominated by the low-frequency behavior of the noise spectral density. 
Expressing the corresponding Lagrange multiplier $\lambda$ through $A_n$, we obtain
\begin{eqnarray}
    S=\frac{n\lambda A_n}{2}\sim  T^{1+\beta-2/n} A_n^{2/n}.
\end{eqnarray}
For $\beta=1-2H$ this scaling coincides with the one in Eq.~(\ref{action_const}). Therefore, for an arbitrary stationary Gaussian process with a spectral density that diverges at zero frequency, the optimal path is expected to be delocalized for $1\leq n \leq 2/(1+\beta)$. For $n>2/(1+\beta)$, a localized solution yields a lesser action and is therefore optimal.

\section{Numerics}
\label{Numerics}

\subsection{$n=2$: Numerical evaluation of the eigenvalue $\lambda$}
\label{Nystrom}
To test our analytical predictions for $n=2$ (see Section \ref{OFM2}), we discretized the continuous linear integral equation (\ref{n=2FE}) into a matrix eigenvalue problem by using numerical integration. Dividing the integration interval $[-T, T]$ into $N$ subintervals, with grid points  $t_j = -T + (2T/N)j, \quad j = 0, 1, 2, \ldots, N$,  one can rewrite Eq. (\ref{n=2FE}) in a matrix form:
\begin{equation}
   X = 2 \lambda h M X,
\end{equation}
where $h=2T/N$ is the lattice step size, $X^T=\begin{pmatrix} x(t_0), & x(t_1), &  \ldots, & x(t_{N-1}), & x(t_N) \end{pmatrix}$ represents the discretized optimal path, and $M_{l,j} = \kappa(t_l- t_j)$ is the covariance matrix. The minimal eigenvalue $\lambda$ of Eq.~(\ref{n=2FE}) corresponds to the maximal eigenvalue of the symmetric covariance matrix $M$, while $X$ plays a role of the associated eigenvector.

Once the maximal eigenvalue $\mu_\text{max}$ of the matrix $M$ is found, we obtain the Lagrange multiplier $\lambda=1/(2 h \mu_\text{max})$ and the action $S(A_2,T)=\lambda A_2$.   Without loss of generality we set $A_2=1$, $\gamma=1/4$, and $D = 1$.  The accuracy of this numerical approach for the lattice step size $h=0.1$ was tested by comparing the numerically found action with the exact analytical action for $H=1/2$, $S(A_2, T)=\frac{\gamma^2 A_2}{4 D}$. For example, for $T=1000$ and $A_2=1$ the  relative error,
$$
\left|\frac{S(A_2,T)-S_{\text{num}}(A_2,T)}{S(A_2,T)}\right|\,,
$$
of the numerically estimated $S_{\text{num}}(A_2,T)=\lambda A_2$ was of the order of $10^{-5}$.

\begin{figure}[ht]
\centering
\includegraphics[clip,width=1.02\textwidth]{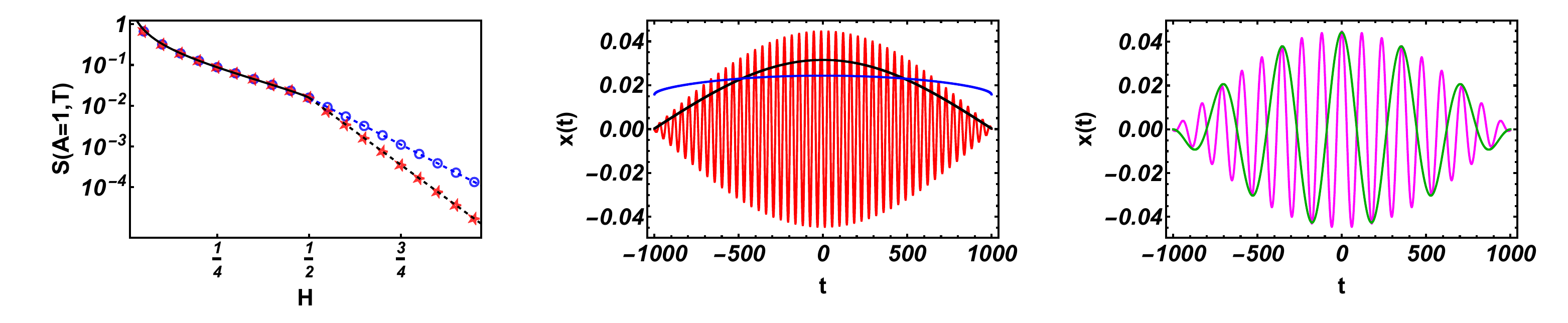}
\caption{Left: Numerically computed actions for $n=2$ (the blue circles correspond to $2T=200$, the red stars correspond to $2T=2000$) are compared with the theoretical prediction (\ref{n=2_act}) (the blue and black dashed lines, respectively). The black solid line represents the action (\ref{Sbound}) for $H\leq 1/2$. Center: the optimal paths $x(t)$ for $H=1/4$ (red), $H=1/2$ (black), and $H=3/4$ (blue). Right:  the optimal paths $x(t)$ for $2T=2000$ and two values of $H$: $H=0.457$ (magenta) and $0.495$ (green), for which $\omega_{\text{max}}(0.457)\simeq 3 \omega_{\text{max}}(0.495)$. In all three panels we set  $A_2=1$, $\gamma=1/4$ and $D=1$.}
\label{fig_EV}
\end{figure}

The numerical results for the  actions $S(A_2=1,T)$ for $n=2$ and two integration times $2T=200$ and $2T=2000$ are shown in the left panel of Fig.~\ref{fig_EV}. The markers show the numerically computed actions, the solid black line represents the theoretically predicted action~(\ref{Sbound}) for the short-correlated fOU process ($H\leq 1/2$), while the two dashed lines represent the prediction~(\ref{n=2_act}) for the long-correlated fOU process ($H>1/2$). One can see that the action~(\ref{Sbound}), obtained by extending the integration limits to infinity, is in excellent agreement with the numerical results. For $H>1/2$ the predicted action (\ref{n=2_act}) is also  in agreement with the numerically computed action.

The central panel of Fig.~\ref{fig_EV} shows the numerical optimal paths $x(t)$, computed as the eigenvectors of the covariance matrix $M$ corresponding to its maximal eigenvalues, for $2T=2000$ and three values of the Hurst exponent $H$. For the long-correlated fOU ($H=3/4$), the optimal path is close to a constant (notice the huge difference between the horizontal and vertical scales), as we assumed in Section \ref{OFM2}.  Because of the long correlations, $x(t)$ remains non-zero well beyond the integration interval $|t|<T$ (not shown).  For $H=1/2$  the optimal path is very close to a constant for $|t|<T$, and it falls off exponentially fast, on the time scale $1/\gamma$, beyond the integration interval (not shown).

For $H<1/2$ the numerical optimal path oscillates at a frequency very close to its theoretically predicted value (\ref{omegamax}). An additional very slow modulation, on the time scale $2T$,  is a finite-size effect, which has only  a negligible effect on the action. Still, for completeness, we calculate in Appendix \ref{appA} finite-size corrections for $H=1/2$, where they can be easily determined analytically.

\subsection{$n>2$: Numerical iterations}
\label{NumSol}
To test our non-rigorous scaling arguments for $n>2$, we  solved the integral equation (\ref{yn}) numerically by iterations, using a straightforward iteration procedure
\begin{equation}
    y_i(t)=\int\limits_{-T}^{T}dt' \kappa(t -t')y_{i-1}^{n-1}(t').
    \label{yn_d}
\end{equation}
Once a numerical solution $y(t)$ was found with a required accuracy, we determined the Lagrange multiplier $\lambda$ from Eq.~(\ref{constraint}).  Then we computed the optimal path $x(t)=(n\lambda)^{-1/(n-2)}y(t)$ in the original variables and, using Eq.~(\ref{act-labmda}), the action $S(A_n,T)$. We set $A_n=1$, $\gamma=1/4$, and $D = 1$ without loss of generality [the areas $A_n$ can be set to $1$ because of the universal OFM scaling of the optimal solution with $A_n$, see Eq.~(\ref{scaling-action})]. Some additional details of the numerical implementation of the iteration process are presented in Appendix \ref{appB}.

\begin{figure}[ht]
\centering
\includegraphics[clip,width=0.8\textwidth]{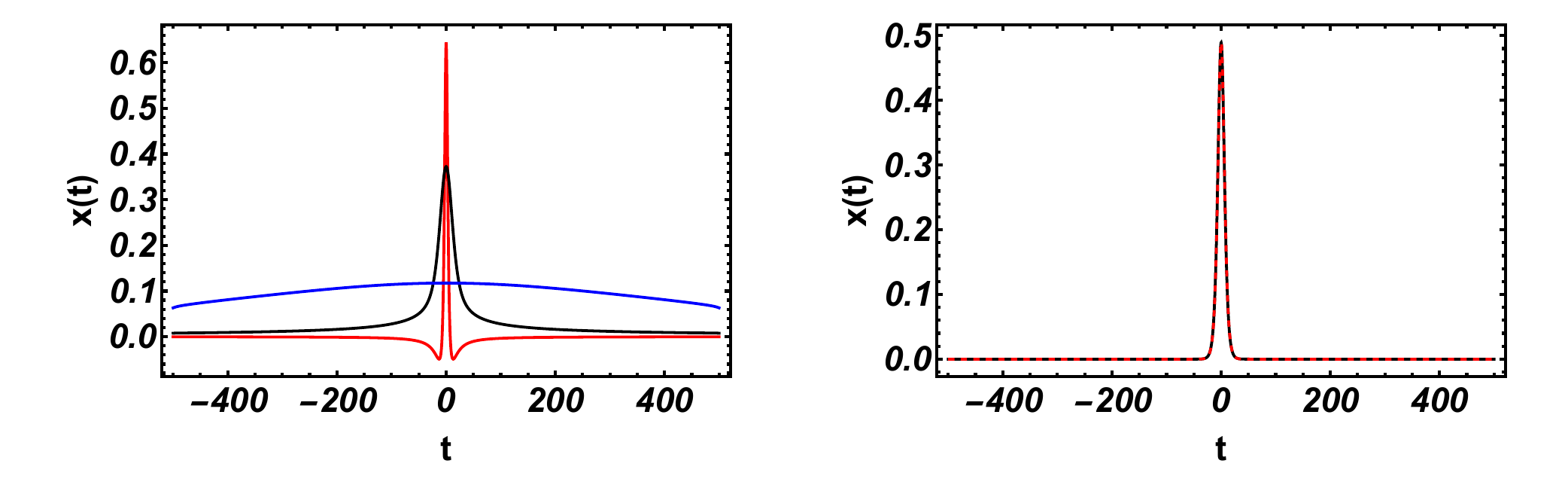}
\caption{Numerically computed optimal paths $x(t)$ for $n=3$, conditioned on $A_3=1$. The parameters are $2T=1000$ and $\gamma=1/4$.
Left: $x(t)$ for the Hurst exponents $H=1/4$ (red), $H=5/8$ (black) and $H=3/4$ (blue). Right:  numerical solution for $H=1/2$ (red dashed line) vs. exact analytical optimal path at $T\to \infty$, Eq.~(\ref{OUx(t)}).}
\label{fig_y}
\end{figure}

The numerically computed optimal paths $x(t)$ for $n=3$ and three values of $H$  are shown in Fig. \ref{fig_y}. To determine the scaling behavior of these optimal trajectories with $T$, see Eq.~(\ref{inter_x}), we plotted their amplitude $x(t=0)$ and full width at half-maximum $\sigma$ vs. $T$. For $H<2/3$ the optimal paths are strongly localized near $t=0$ with $\sigma$ independent of $T$, so that $\chi=0$, as it has been predicted in Section~\ref{SectionHM}.  For $H>3/4$ fully delocalized optimal paths emerge with $\sigma \propto T$, that is $\chi=1$, again as has been predicted. Furthermore, for $H>3/4$ the optimal solutions are almost constant, which justifies the use of the constant solution for the calculation of the action in Eq.~(\ref{action_const}). The right panel of Fig. \ref{fig_y} tests our numerical iteration method by comparing the numerical optimal path for $H=1/2$ (the dashed red line) with the exact solution (\ref{OUx(t)}) at $T\to \infty$ (the black line).

\begin{figure}[ht]
\centering
\includegraphics[clip,width=1.02\textwidth]{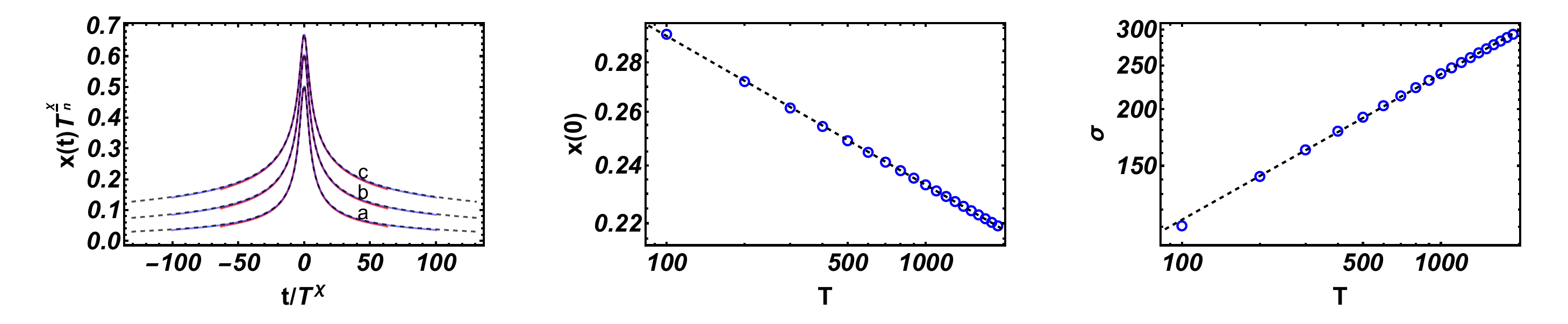}
\caption{Left: Rescaled numerically computed optimal paths $T^{\chi/n}x(t)$, conditioned on $A_n=1$, are plotted against the rescaled time $t/T^\chi$ with $\chi=1/3$ for $n=3$ (a), $n=4$ (b), and $n=5$ (c). For each value of $n$, three trajectories are shown, corresponding to $2T=1000$ (red), $2T=2000$ (blue) and $2T=3000$ (black dashed). Center: the amplitude $x(0)$ of the optimal trajectory for $n=3$ vs. $T$ (blue circles). The black dashed line shows the power-law fit $0.455\times  T^{-0.097}$. Right: the width $\sigma$ of the optimal trajectory, conditioned on $A_n=1$, for $n=3$ vs. $T$ (blue circles). The black dashed line shows the power-law fit $25.92 \times T^{0.322}$. We set $\gamma=1/4$.} 
\label{fig_border}
\end{figure}

Figure~\ref{fig_border}  presents our numerical results for the optimal paths on the borderline $H=1-1/n$ of the phase diagram. The central and right panels show  power-law fits of the amplitude $x(0)$, and of the characteristic width $\sigma$, respectively,  vs. $T$ for $n=3$. The fit gives $x(0)\sim T^{-0.097}$ for the amplitude and $\sigma\sim T^{0.322}$ for the width. These results are in a fair agreement with Eq.~(\ref{inter_x}), and they suggest that $\chi=1/3$. This value of $\chi$ appears to be independent of $n$, as we checked for $n=3,4$ and $5$. To highlight the self-similar structure (\ref{inter_x}) of the optimal paths on the borderline $H=1-1/n$, the left panel of Fig.~\ref{fig_border} shows, for $n=3,4$, and $5$, the rescaled optimal paths  $T^{\chi/n} x(t)$ versus the rescaled time $t/T^\chi$ for three different values of the integration time $T$. As one can see, for each of these values of $n$, the three rescaled paths collapse onto a single curve, confirming the underlying self-similarity.

\begin{figure}[ht]
\centering
\includegraphics[clip,width=1.02\textwidth]{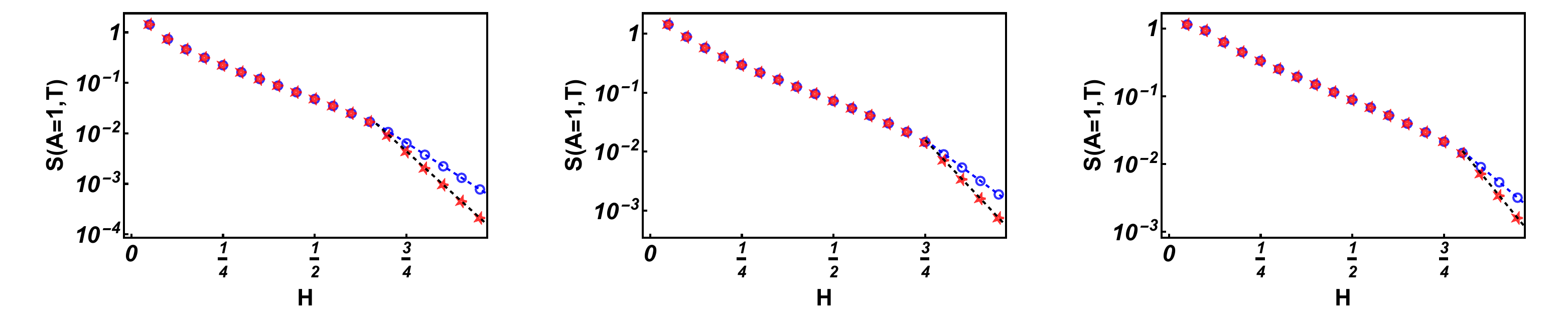}
\caption{Numerically computed actions $S(A_n=1,T)$  for  $n=3$ (left), $n=4$ (center) and $n=5$ (right). The blue circles correspond to $2T=200$, the red stars correspond to $2T=2000$. The blue and black dashed lines show the action (\ref{action_const}), corresponding to the constant solution, for $T=100$ and $T=1000$, respectively. As we predicted, at $H=1-1/n$, with the corresponding $n$,  there is a sharp boundary between the $T$-independent and $T$-dependent actions, originating from the strongly localized and delocalized optimal paths, respectively.}
\label{fig_S}
\end{figure}

Our numerical results for the actions $S(A_n=1,T)$ for $n=3$, $4$, and $5$ and two integration times $2T=200$ and $2T=2000$, are shown in Fig.~\ref{fig_S}. Also shown are predictions from Eq.~(\ref{action_const}). For $H<1-1/n$, the actions for the different $T$ coincide, fully confirming the existence of a localized solution  and the resulting scaling dependence (\ref{act-local}). For $H>1-1/n$, the numerically computed actions depend on the integration time and, as one can see from Fig.~\ref{fig_S}, are in good agreement with Eq.~(\ref{action_const}).

\begin{figure}[ht]
\centering
\includegraphics[clip,width=0.35\textwidth]{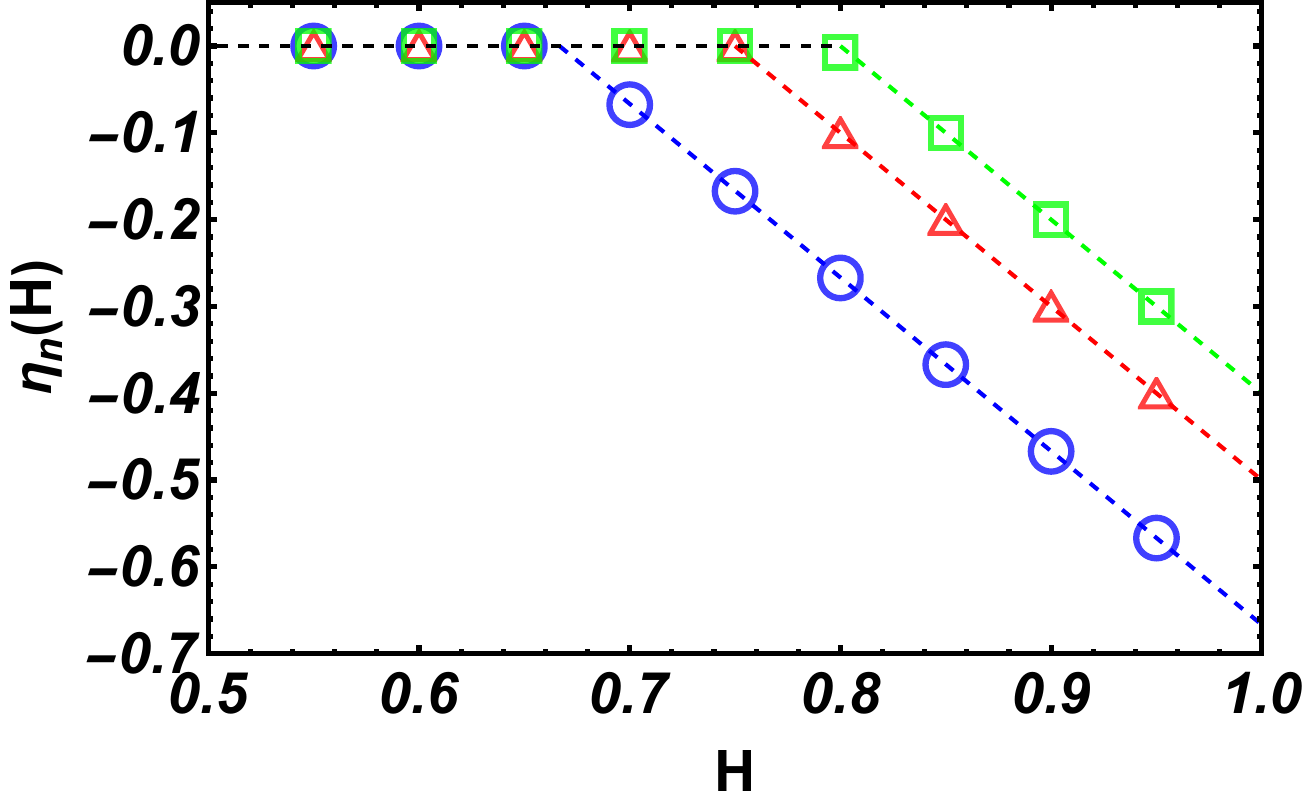}
\caption{The exponent $\eta_n(H)$, entering the scaling relation $S(A_n=1,T)\sim T^{\eta_n(H)}$, see Eq.~(\ref{A_T}), for $n=3$ (blue circles), $n=4$ (red triangles) and $n=5$ (green squares). The dashed lines of the corresponding colors depict the lines $2-2H-2/n$ [see Eq.~(\ref{A_T})] for $n=3$, $4$ and $5$. The black dashed line corresponds to $\eta_n(H)=0$.}
\label{fig_Scaling}
\end{figure}

Repeating the computations of the numerical actions for different $T$ and  $H$, we extracted the scaling dependence $S(A_n=1,T)\sim T^{\eta_n(H)}$ of the numerically estimated action. The scaling exponents $\eta_n(H)$ for $n=3$, $4$, and $5$  are shown in Figure \ref{fig_Scaling}. As one can see, the agreement between the predicted scaling behaviors~(\ref{A_T}) and the numerical solutions is excellent in all the three cases.

\subsection{Wang-Landau simulations}

As an independent verification of our results for the large-$A_n$ tail of the distribution $\mathcal{P}(A_n, T)$, presented in the previous sections, we performed large-deviation Monte-Carlo simulations. We implemented the Wang-Landau (WL) algorithm \cite{WL1, WL2} to sample discretized configurations of the fOU process $x(t)$ for  $n=3$. The simulations were performed by discretizing the process over a large but finite domain, represented as $\Vec{X} = (X(1), X(2), \ldots, X(L))$, where $L \gg 1$ and the lattice step is $\Delta t = 1$. Since sampling a  Gaussian stationary process directly is computationally expensive, we employed the Circulant Embedding Method (CEM)
\cite{Chan1994,Dietrich1997},  which allows sample generation in $O(L \log L)$ time.

The WL algorithm   efficiently estimates the density of states $\mathcal{P}(A_n,T) \sim  \exp[-S(A_n,T)]$, where the area $A_n$ [see Eq.~(\ref{EmpMean})] of the discretized process $\Vec{X}$ is approximated, for $n=3$, by the discrete sum
\begin{equation}
    A_n=\sum\limits_{k=L/2-T}^{L/2+T} X^3(t_k).
\end{equation}
For more details on the simulations, see Ref.  \cite{VLM2024}, where  the CEM and the  WL algorithm were employed to study the barrier height  distribution of a homogeneous Gaussian disorder potential.

We used a regular lattice of length $L = 10^3$ and three integration times: $2T=250,\, 500$, and $750$. The parameter $\gamma=0.05$, which determines the characteristic time scale $\sim\gamma^{-1}$ of the covariance function (\ref{cov}), was chosen to be sufficiently small so as to accurately approximate a continuous stationary Gaussian process on a discrete lattice with unit lattice step. For convenience, the diffusion coefficient was chosen $D=\frac{\gamma^{2H}}{\Gamma(2H+1)}$.

\begin{figure}[ht]
\centering
\includegraphics[clip,width=0.7\textwidth]{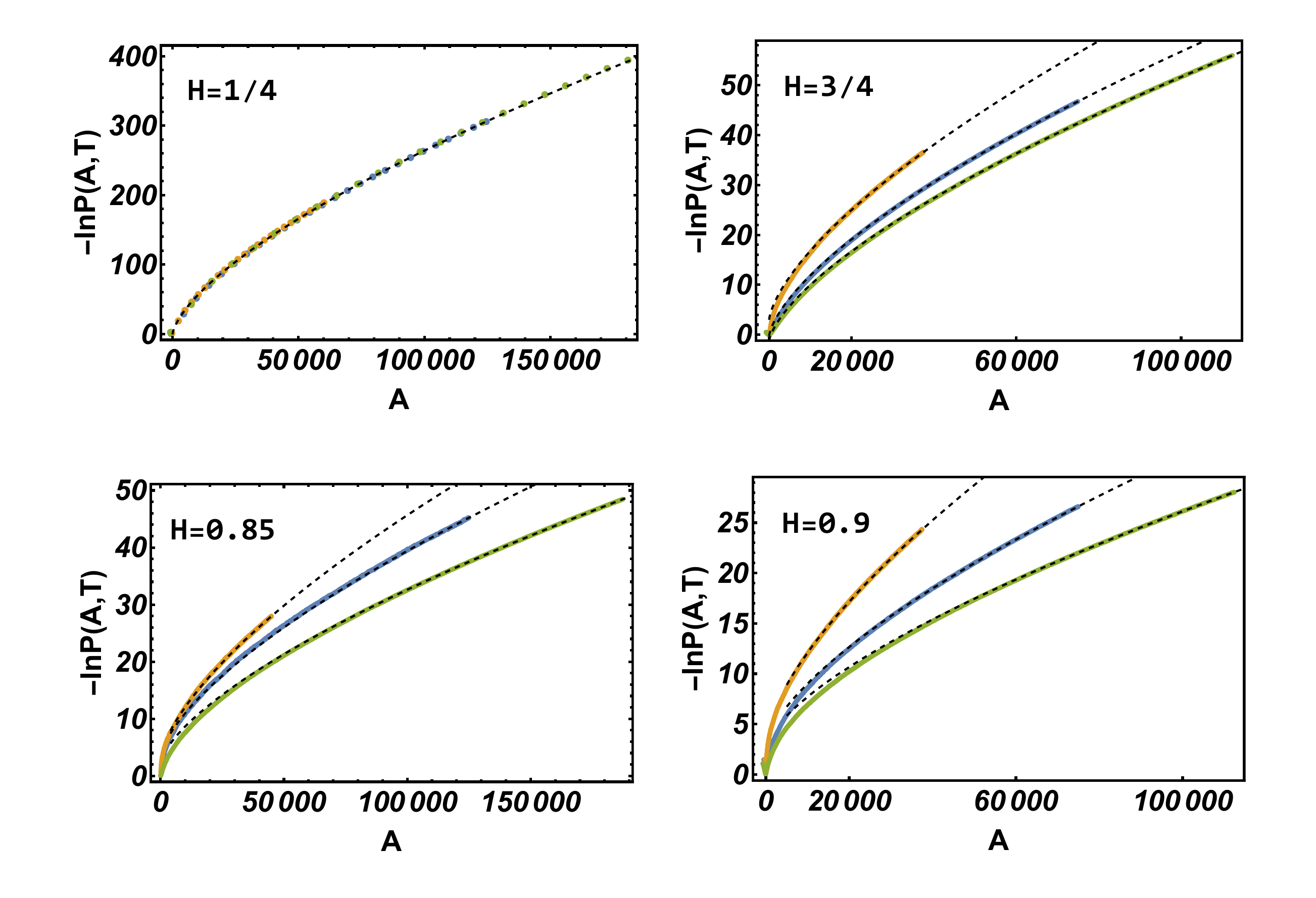}
\caption{The large-$A_n$ behavior of the action, $-\ln \mathcal{P}(A_n,T)$, as measured in the WL simulations for set of $H$ (different panels) and $2T =250$ (orange), $500$ (blue) and $750$ (green) .  The asymptotic behavior~(\ref{Sapp}) is shown in all cases by the dashed black lines.}
\label{fig2}
\end{figure}

The simulation results for the action  $S(A_n,T)=-\ln \mathcal{P}(A_n, T)$ are presented in  Fig. \ref{fig2} alongside with the predicted asymptotic behavior
\begin{eqnarray}
    S(A_n\to \infty, T\to \infty)\simeq c A_n^{2/3}
    \label{Sapp}
\end{eqnarray}
with fitted coefficients $c=c(T,D, H,\gamma,n)$. According to Eq.~(\ref{A_T}), for $H<2/3$, the actions $S(A_n, T)$, and therefore the fitted coefficients $c$, should be independent of $T$, and this property is reproduced in the simulations for $H=1/4$. The situation is the opposite for $H>2/3$, as represented by $H=3/4$, $0.85$ and $0.90$, where the actions and the coefficients $c$ strongly depend on the integration time $T$.

\begin{figure}[ht]
\centering
\includegraphics[clip,width=0.8\textwidth]{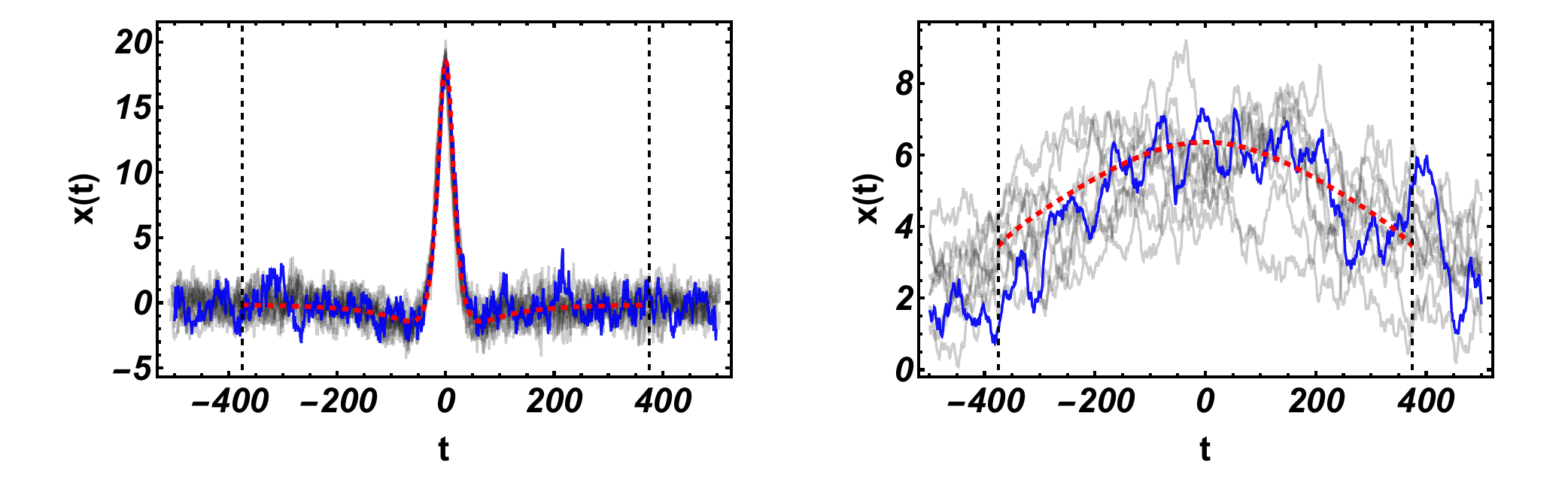}
\caption{$10$ realizations of the fOU process $x(t)$  corresponding to $n=3$ and $A_3=112 500$ for $H=1/4$ (left panel) and $H=3/4$ (right panel), as obtained in the WL simulations.}
\label{fig2a}
\end{figure}

Several realizations of the fOU process $x(t)$, conditioned on a large $A_n$ and sampled in our WL simulations, are presented in Figure \ref{fig2a}. The red dashed lines represent the numerical optimal paths computed using the iteration method, as explained in Section \ref{NumSol}. As one can see, for $H=1/4$ the realizations are very close to each other, and to the predicted optimal path, in the region of the strong localization, which dominates the action. For $H=3/4$ the realizations look ``more random". Note, however, a huge difference between the vertical and horizontal scales in the two cases. For this value of $H$ the realizations essentially linger close to zero, but they still follow the
predicted optimal behavior ``on the average". The spread of different sample trajectories around the optimal path for $H=1/4$ and $H=3/4$ is actually comparable.

\begin{figure}[ht]
\centering
\includegraphics[clip,width=0.35\textwidth]{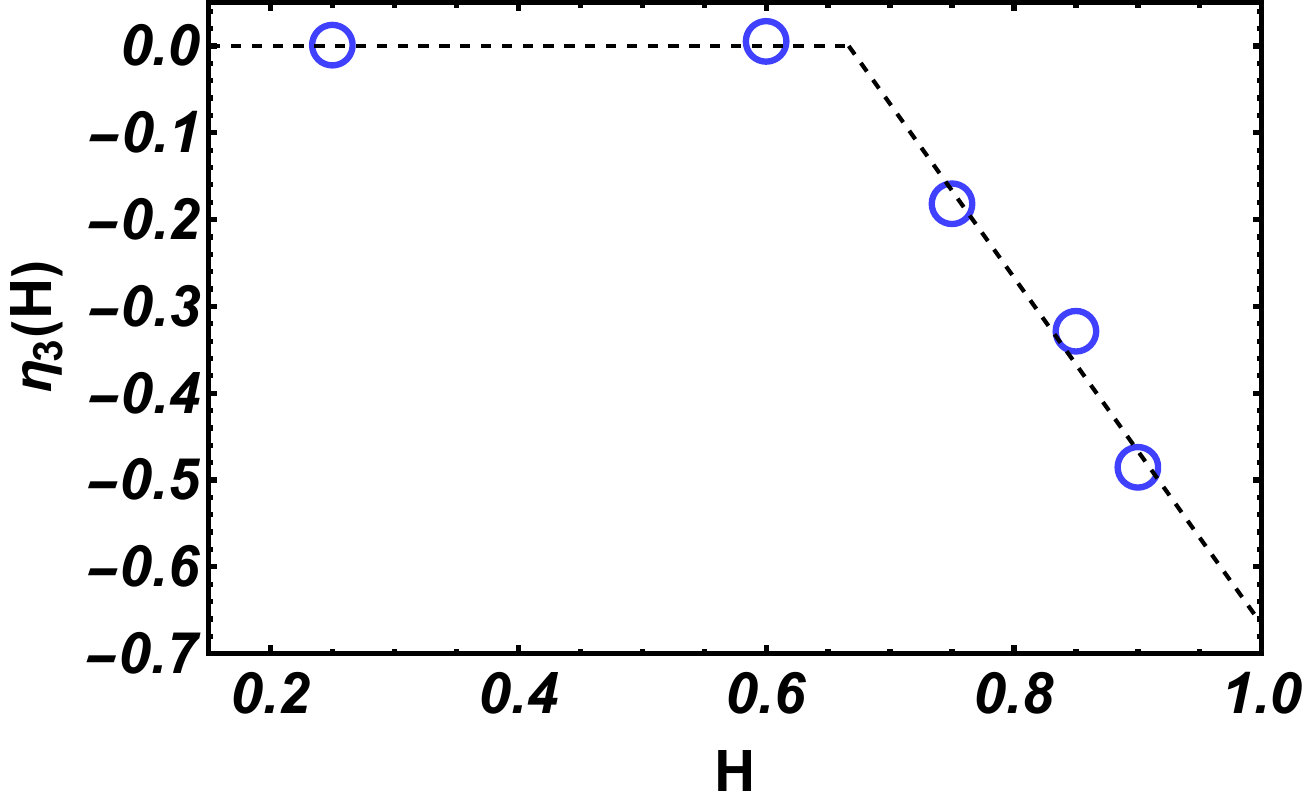}
\caption{Scaling behavior of the action $S(A_3,T)\propto T^{\eta_3(H)}$, see Eq.~(\ref{Sapp}), as a function of $H$. Blue circles: simulations. Black dashed line: $\eta_3(H)=0$ for $H\leq 2/3$ and  $\eta_3(H) = 2-2H-2/3$ for $H>2/3$, as predicted in  Eq.~(\ref{A_T}). }
\label{fig3}
\end{figure}

To extract the scaling dependence of the simulated action $-\ln \mathcal{P}(A_n, T)$ on $T$ we examined the behavior of the fitted coefficients $c$ for different $T$. The results, for different $H$, are shown in Fig. (\ref{fig3}). As one can see, they coincide with the scaling behavior presented in  Fig. \ref{fig_Scaling} and faithfully reproduce our predictions~(\ref{A_T}).

\newpage
\section{Summary and Discussion}

We extended the previous studies of large deviations of time-integrated observables (\ref{EmpMean}) of
(in general, non-Markovian) stationary Gaussian processes to the situations where the spectral density of the process either vanishes
or diverges at zero frequency. Focusing on the important example of the fOU process with the Hurst exponent $0<H<1$ and  applying the OFM,  we have found that, in all cases,
the action $S=-\ln \mathcal{P}(A_n,T)$  scales as $A_n^{2/n}$ thus showing a remarkable universality.  The scaling behavior of the action with the integration time $T$, however, is quite diverse, and we uncovered a nontrivial phase diagram of these behaviors on the $(H,n)$ plane, see Eqs. (\ref{1}) - (\ref{3}) and Fig. \ref{fig_PhaseDiagram}. These different behaviors result from the differences in the character of the corresponding optimal paths of the conditioned process. We have observed two main types of the optimal paths: strongly localized paths and completely delocalized ones. The ensuing competition between them in terms of their action determines the phase diagram of the scaling behaviors. However, along the borderline $H=1-1/n$  on the $(H,n)$ plane a more subtle behavior is observed, as the optimal path is ``intermediately localized" on a time scale which grows with $T$ sublinearly, apparently as $T^{1/3}$. We verified our theoretical and numerical OFM predictions in large-deviation simulations of the conditioned fOU process, where we were able to probe  probability densities as small as $10^{-170}$.

One important take-home message from these results is that, for long-correlated processes (as exemplified by the fOU process with $H>1/2$), the strongly localized optimal paths for $n>2$ can give way to delocalized optimal paths, with dramatic consequences for the scaling behavior of the action with the integration time $T$. This happens, however, only for ``sufficiently fat" correlation tails, as exemplified by the condition $H>1-1/n$ for the fOU process.

It has been found recently for Markov processes \cite{Smith2022,Smith2024} that  different scaling behaviors  of the typical fluctuations and of the large deviations of $A_n$ may lead to a dynamical phase transition  (DPT) -- a singularity in the large-deviation function $S(A_n,T)$ at $n>2$. It would be very interesting to explore analogous  dynamical phase transitions here.

\section*{Acknowledgments}
We are very grateful to Naftali R. Smith for useful comments. The authors  were supported by the Israel Science Foundation (Grant No. 1499/20).

\appendix

\section{Finite-size effects for $H=1/2$ and $n=2$}
\label{appA}

For $H=1/2$ the OFM problem reduces to a minimization of the local action functional of the OU process, which results in an ordinary Euler-Lagrange equation. For $n=2$ the latter equation is linear:
\begin{equation}
    \ddot{x}(t)-\gamma^2 x(t)=-4D\lambda x(t)\big[\theta(t+T)-\theta(t-T)\big].
\end{equation}
The general solution, symmetric with respect to the time reversal, reads
\begin{equation}
    x(t)=\begin{cases}
        F \cos\left(\sqrt{4D\lambda-\gamma^2}t\right),\qquad &\text{for $|t|\leq T$,}\\
        E \exp[-\gamma |t|],\qquad &\text{for $|t|> T$}.
    \end{cases}
    \label{gensol}
\end{equation}
The constants $F$ and $E$ and the Lagrange multiplier $\lambda$ can be determined from the area constraint (\ref{EmpMean}) and the continuity of $x(t)$ and $\dot{x}(t)$ at $t=\pm T$.
In the limit of $T\to \infty$ one obtains $\lambda=\gamma^2/(4D)$, and the solution  on $|t|<T$ becomes constant: $x(t)=\sqrt{A_2/2T}$, leading to the action, described by the second line of Eq.~(\ref{NT}).

For finite $T$ the continuity of $x$ and $\dot{x}$ yields a transcendental equation for the Lagrange multiplier $\lambda$, which can be written as follows:
\begin{equation}
    z \tan (z)=\gamma T\,,
    \label{eq3}
\end{equation}
where $z=T \sqrt{4D\lambda-\gamma^2}$.
At large $\gamma T$ one can expand the l.h.s. of Eq.~(\ref{eq3}) in a vicinity of the $z=\pi/2-\epsilon$, where $\epsilon \ll 1$. We obtain
\begin{equation}
\frac{\pi}{2\epsilon}-1\simeq \gamma T.
\end{equation}
The solution of this equation in the leading order is
\begin{equation}
    \epsilon\simeq \frac{\pi}{2\gamma T}\,.
\end{equation}
Expressing  $\lambda$ through $\epsilon$ and keeping only first subleading term, we obtain
\begin{equation}
    \lambda=\frac{\gamma^2}{4D}+\frac{\pi^2}{16 D T^2}+\mathcal{O}\left(\frac{1}{T^3}\right)\,.
\end{equation}
The optimal path, satisfying the area constraint, is
\begin{equation}
    x(|t|<T)\simeq\sqrt{\frac{A_2}{T}} \cos\left(\frac{\pi t}{2 T}\right).
\end{equation}
The action, with an account of the leading finite-size correction, is the following:
\begin{equation}\label{correctedS}
S(A_2,T)=\lambda A_2\simeq \frac{\gamma^2 A_2}{4D}+\frac{\pi^2A_2}{16 D T^2}+\mathcal{O}\left(\frac{1}{T^3}\right)\,.
\end{equation}
The leading correction is $\mathcal{O}(T^{-2})$, and it is negligible in the limit of large $T$.

\section{Numerical iteration algorithm}
\label{appB}

To apply the iteration scheme (\ref{yn_d}), we discretize the integral in a standard way, by subdividing the interval $[-T, T]$ into $N$ subintervals, with grid points  $t_j = -T + j (2T/N), \quad j = 0, 1, 2, \ldots, N$, where $N=40T$. Then the integral is evaluated by using a quadrature rule. Using this discretization 
in Eq.~(\ref{yn_d}) we obtain the following system of equations for the values of  $y(t)$  at the grid points $t_j$, where \( j = 0, 1, \ldots, N \):
\begin{equation}
    y_i(t_j)=\sum\limits_{l=0}^{N} h \kappa(t_l -t_j)y_{i-1}^{n-1}(t_l).
    \label{discr}
\end{equation}
where $h=2T/N$. To simplify the iteration process we rewrite this set of equations  in a matrix form. Let $Y_i$ represent the vector of values of the approximate solution at the grid points at $i$-th iteration and  $P_{n}(Y_i)$ represent the vector of Hadamard power of $Y_i$:
\begin{equation}
 Y_i = \begin{pmatrix} y_i(t_0) \\ y_i(t_1) \\ \vdots \\ y_i(t_N) \end{pmatrix} \quad\text{and}\quad P_{n}(Y_i) = \begin{pmatrix} y^n_i(t_0)) \\ y^n_i(t_1)) \\ \vdots \\ y^n_i(t_N)) \end{pmatrix}.
\end{equation}
We define the covariance matrix $M$ as
\begin{equation}
M_{l,j} = \kappa(t_l- t_j),
\end{equation}
Then, the matrix form of the discretized system (\ref{discr}) is
\begin{equation}
   Y_{i} = h M P_{n-1}(Y_{i-1}).
\end{equation}
We start the iteration process with a flat initial guess $Y_0$,
\begin{equation}
Y^T_0 = \begin{pmatrix} 1, & 1, &  \ldots, & 1, & 1 \end{pmatrix}\,.
\end{equation}
We iterate until the relative error of the normalized solutions $\{y_{i}(t_j)/||y_i||\}$ becomes sufficiently small:
\begin{equation}
      \underset{j}{\text{max}}\left|\frac{y_{i}(t_j)/||y_i||-y_{i-1}(t_j)/||y_{i-1}||}{y_{i}(t_j)/||y_i||}\right|<10^{-8},
\end{equation}
where $||y_i||^2=h\sum\limits_{j=0}^{N}y_{i}^{2}(t_j).$
The convergence of the iteration process typically requires about $20$ iterations to reach the solution on the interval $(-T,T)$, where $T\sim 10^3$.

Once an approximate solution $y(t)$ is found, it can be used to evaluate the Lagrange multiplier $\lambda$, the optimal path $x(t)=(n\lambda)^{-1/(n-2)} y(t)$, and the action via the following equations:
\begin{equation}
\begin{cases}
    n\lambda=A_n^{\frac{2-n}{n}}\left(\sum\limits_{j=0}^{N} hy^{n}(t_j)\right)^{\frac{n-2}{n}}, \\
    x(t)=A_n^{1/n}\left(\sum\limits_{j=0}^{N} hy^{n}(t_j)\right)^{-1/n}y(t), \\
    S(A_n,T)=\frac{n\lambda A_n}{2}=\frac{1}{2}\left(\sum\limits_{j=0}^{N} h\tilde{y^{n}}(t_j)\right)^{\frac{n-2}{n}}A_n^{\frac{2}{n}}.
    \label{Num}
\end{cases}
\end{equation}

To verify the accuracy of the iteration procedure, we compared the iterative solution  with the analytical solution for $H=1/2$, see Fig. \ref{fig_y}. The relative error,
\begin{equation}\label{relative}
 \left|\frac{S(A_3,T)-S_{\text{num}}(A_3,T)}{S(A_3,T)}\right|\,,
\end{equation}
of the numerical action $S_{\text{num}}(A_3=1,T)$, see Eq.~(\ref{Num}), compared with the exact action
\begin{equation}\label{exactS3}
S(A_3=1,T)=\left(\frac{9}{10}\right)^{1/3}\frac{\gamma^{5/3}}{2 D}(2T)^{-1/3}
\end{equation}
(see Eq. (15)  in the first paper of  Ref. \cite{NT2018}), was of the order of $10^{-5}$.


\begin{thebibliography}{99}
\bibitem{dimensions} Here are the relevant dimensional quantities with their units, denoted by square brackets: $[T]=\text{time}$,  $[\gamma]=\text{time}^{-1}$, $[D]=\text{length}^2/\text{time}^{2H}$, and $[A_n]=[a_n][T]=\text{length}^n\times\text{time}$. One can construct from them two independent dimensionless parameters \cite{Barenblatt}. The additional two dimensionless parameters are $H$ and $n$. Finally, $\mathcal{P}(A_n)$ has the units of $\text{length}^{-n}\times \text{time}^{-1}$. These arguments lead to the exact scaling behavior presented
    in Eq.~(\ref{scaling}).
\end{thebibliography}
\end{document}